# Unconventional band structure via combined molecular orbital and lattice symmetries in a surface-confined metallated graphdiyne sheet

Ignacio Piquero-Zulaica[1,†,*], Wenqi Hu[2,†], Ari Paavo Seitsonen[3,†], Felix Haag[1], Johannes Küchle[1], Francesco Allegretti[1], Yuanhao Lyu[2], Lan Chen[2], Kehui Wu[2], Zakaria M. Abd El-Fattah[4], Ethem Aktürk[5], Svetlana Klyatskaya[6], Mario Ruben[6,7], Matthias Muntwiler[8], Johannes V. Barth[1,*] and Yi-Qi Zhang[2,1*]

[1]*Physics Department E20, Technical University of Munich, D-85748 Garching, Germany*

[2]*Institute of Physics, Chinese Academy of Sciences, 100190 Beijing, China*

[3]*Département de Chimie, École Normale Supérieure, 24 rue Lhomond, F-75005 Paris, France*

[4]*Physics Department, Faculty of Science, Al-Azhar University, Nasr City, E-11884, Cairo, Egypt*

[5]*Department of Physics, Adnan Menderes University, 09100 Aydin, Turkey*

[6]*Institute of Nanotechnology, Karlsruhe Institute of Technology, 76344 Eggenstein-Leopoldshafen, Germany*

[7]*IPCMS-CNRS, Université de Strasbourg, 23 rue de Loess, 67034 Strasbourg, France*

[8]*Paul Scherrer Institute, Forschungsstrasse 111, 5232 Villigen PSI, Switzerland*

†These authors contributed equally.
*Email: ge46biq@mytum.de; jvb@tum.de; yiqi.zhang@iphy.ac.cn

## Abstract

Graphyne (GY) and graphdiyne (GDY)-based materials represent an intriguing class of two-dimensional (2D) carbon-rich networks with tunable structures and properties surpassing those of graphene. However, the challenge of fabricating atomically well-defined crystalline GY/GDY-based systems largely hinders detailed electronic structure characterizations. Here, we report the emergence of an unconventional band structure in mesoscopically regular (~1 μm) metallated GDY sheets featuring a honeycomb lattice on Ag(111) substrates. Employing complementary scanning tunnelling and angle-resolved photoemission spectroscopies, electronic band formation with a gap of 2.5 eV is rigorously determined in agreement with real-space electronic characteristics. Extensive density functional theory calculations corroborate our observations as well as recent theoretical predictions that doubly degenerate frontier molecular orbitals on a honeycomb lattice give rise to flat, Dirac and Kagome bands close to Fermi level. These results illustrate the tremendous potential of engineering novel band structures via molecular orbital and lattice symmetries in atomically precise 2D carbon scaffolds.

Synthesizing and characterizing π-conjugated two-dimensional (2D) polymers with atomic thickness and covalently bonded periodic structures has recently been pushed to the forefront of research in chemistry and physics[1-4], concomitant with the rapidly advancing exploration of 2D materials[5-7]. Through a suitable choice of molecular building blocks and linking motifs, crystalline 2D organic polymers or hybrid metal-organic sheets can be rationally designed to display principal electronic properties like π-orbital overlapping, dispersive bands as well as adjustable band gap[8,9], which are essential for organic electronics applications[10,11]. In particular, theoretical modelling shows that a proper combination of frontier molecular orbital (MO) and lattice symmetries provides a new avenue towards realizing unconventional band structures in these materials, such as flat, Dirac and Kagome bands[12-15], which can be further tailored to give rise to novel topological or many-body states[13,14,16].

Among various types of π-conjugated 2D polymers, a special class of carbon-based networks distinct from graphene, known as graphyne (GY)[17] and graphdiyne (GDY)[18] is of particular interest[3,19]. Acetylenic precursors containing carbon-carbon triple bonds (–C≡C–) enable a combination of *sp*- and *sp*$^2$-hybridized carbon atoms to create a great variety of GY/GDY-related structures[20], presenting remarkable tunability in their physical and chemical properties[19,21]. However, synthesizing crystalline GY/GDY sheets as well as their analogues turned out to be particularly challenging[19], which hampers in-depth electronic structure characterizations. The past fifteen years witnessed the rapid growth in the field of on-surface synthesis in ultra-high vacuum (UHV)[22,23], giving new access to synthesize 2D polymers via the bottom-up approach[24], and many exemplary carbon-based architectures have been achieved[25-29]. Utilizing a terminal alkyne derivative 1,3,5-tris(4-ethynylphenyl)benzene (Ext-TEB)[30] and a highly chemoselective gas-mediated on-surface reaction protocol, we recently fabricated a highly regular mesoscale (~ 1 μm) organometallic monolayer[31], formally representing an Ag-metallated graphdiyne (Ag-GDY) analogue. Therefore, it provides an ideal platform to scrutinize the electronic properties of atomically well-defined metallated GDY sheets.

In this work, employing scanning tunnelling microscopy and spectroscopy (STM/STS), angle-resolved photoelectron spectroscopy (ARPES) and density functional theory (DFT), we demonstrate that the intrinsic electronic structure of Ag-GDY corresponds to a hole-doped semiconductor with an unconventional band structure. Notably, both the conduction band minimum and the valence band maximum comprise nontrivial flat bands, resulting from the combined molecular orbital and lattice symmetries. The band formation together with a gap of ≈ 2.5 eV is rigorously determined via STS and ARPES characterization, whereby a charge transfer from the substrate to the network is induced by mild electronic hybridization. DFT calculations corroborate the experimental results and further reveal that employing an insulating buffer layer may restore the Ag-GDY intrinsic band structure with a flat band lying near the Fermi level, which is important for tuning correlated electronic behaviour in such systems.

**Results**

**Emerging novel band structure in freestanding Ag-GDY.** The Ag-GDY network (cf. chemical scheme in right panel of Fig. 1b) presents unique structural characteristics, featuring a honeycomb lattice occupied by Ext-TEB molecules and a Kagome lattice spanned by alkynyl-Ag atoms (cf. highlighted grids in Fig. 1b, left panel). Therefore, the Ag-GDY network is expected to harbour dispersive bands as Ext-TEB is aromatic[32] and the Ag-intercalated butadiyne bridge (–C≡C–Ag–C≡C–) has covalent character[33]. As a first step, the electronic band structure of the freestanding layer with a rhombic unit cell containing two Ext-TEB molecules and three silver atoms was calculated (cf. Fig. 1c and Supplementary Fig. 1). The conduction bands (CBs) of the network start at ~2.4 eV above the Fermi level (Fig. 1c). The first group (CB$_1$) features two Dirac bands sandwiched between two nearly flat bands, whereas the second group ($CB_2$) contains only two Dirac bands. The valence bands (VBs) reach 0.03 eV above the Fermi level, comprising two consecutive sets of Kagome bands (cf. $VB_1$ and $VB_2$ in Fig. 1c). It is unusual that lowest CB and highest VB are both flat bands, and these features are reminiscent of the typical band structure predicted from a honeycomb lattice with doubly degenerate on-site orbitals[12-14].

To further examine the network band structure characteristics, MOs of Ext-TEB were calculated (cf. Fig. 1a and Supplementary Fig. 2). Both lowest unoccupied and highest occupied MOs (LUMO/HOMO) display double degeneracy, related to the three-fold symmetry of the molecule (Fig. 1a). DFT calculations also show that LUMO+1 and HOMO-1 for Ext-TEB are single levels with different energy separation ΔE to LUMO and HOMO (Fig. 1a), respectively.

The band structure of the Ag-GDY network can be interpreted based on a tight-binding model, as outlined by Ni et al[13], employing a three-orbital (σ, $π_x$, $π_y$) basis on a honeycomb lattice, which results in six bands. If the energy separation ΔE between the degenerate ($π_x$, $π_y$)-MOs and the single σ-MO is large (ΔE = 0.4 eV; red box in Fig. 1a), the former yields four bands with exact characteristics of $CB_1$ (Fig. 1c) and the latter produces two Dirac bands ($CB_2$). Alternatively, a small ΔE (ΔE = 0.23 eV; blue box in Fig. 1a) will cause hybridization between σ and ($π_x$, $π_y$)-MOs, yielding two sets of Kagome bands[13], representing the Ag-GDY valence band edge ($VB_1$ and $VB_2$ in Fig. 1c). To further verify the band composition, isosurfaces of orbital density integrated within the selected band groups were plotted (Fig. 1d). The orbital characters of $CB_1$ and $CB_2$ clearly correspond to LUMO and LUMO+1, respectively, and a similar correlation with HOMO and HOMO-1 exists for the VBs (cf. also Supplementary Fig. 2). Therefore, the emergence of flat, Dirac as well as Kagome bands close to the Fermi level can be rationalized based on the frontier MO and lattice symmetries of the Ag-GDY network. Another intriguing feature of the band structure is that the Fermi level intersects slightly below the flat top VB, which is attributed to the open shell character of the alkynyl-Ag bridge[33-35]. Accordingly, isolated Ag-GDY sheets show hole-doped semiconductor characteristics with leading flat bands enclosing a gap of $E_g \approx 2.4$ eV.

**Extended LDOS characteristics in a single Ag-GDY layer.** High-quality samples are essential for electronic property characterization. The overview STM image in Figure 2a depicts the regular Ag-GDY network prepared on the Ag(111) single crystal surface, covering continuously an area of ~ $500 \times 500$ nm$^2$ without apparent defects (cf. also Supplementary Fig. 3). The long-range order of the atomically thin film is also evidenced by the corresponding exemplary low-energy electron diffraction (LEED) pattern displayed in Fig. 2b. The LEED spots can be modelled via the superposition of three equivalent rectangular lattices along high-symmetry directions of Ag(111)[31] (cf. Fig. 2c and Supplementary Fig. 4). For practical purposes, we also prepared Ag-GDY networks on inexpensive Ag(111)/mica. Notably, the atomically straight steps along high-symmetry directions favour continuous network growth across adjacent terraces (see Fig. 2d and Supplementary Fig. 5). This single layer continuity resembles the main feature of covalently bonded 2D materials (e.g., *h*-BN or graphene sheets)[36,37]. On both substrates the network unit cell is rectangular, with parameters $a_1 = 63.6$ Å, $a_2 = 35.0$ Å (cf. Fig. 2f and Supplementary Fig. 7).

To unravel band formation in Ag-GDY, site-dependent differential conductance (d*I*/d*V* or STS) spectra as well as bias-dependent d*I*/d*V* maps were obtained, focusing on a six-membered ring as basic constituent of the regular network (cf. Fig. 3c). Hereby care must be taken to differentiate the appearance of the well-known quantum confinement effects imposed by the nanoporous lattice[38] on the quasi-2D electron gas provided by the Ag(111) surface state (SS) from the electronic properties of the Ag-GDY sheet.

Fig. 3a depicts point spectra recorded on Ag surface (black and grey curves), Ext-TEB molecules (red) and alkynyl-Ag atoms (blue) in a bias range from -0.2 to 1.2 V. The data taken at the pore centre (black curve) displays three main peaks at 81, 328 and 677 mV, which correspond to the first, second and fourth (n=1, 2, 4) resonances of confined substrate Shockley-SS electrons in the nanocavity[38] (not related to the 2D polymer). The solid red curve in Fig. 3a is averaged over six molecular sites in the hexagonal unit (red mark in Fig. 3c). Genuine band formation is firstly evidenced by the fact that the d$I$/d$V$ spectra taken at six molecular sites are nearly identical (cf. dimmed red curves in Fig. 3a), whereas this conformity was absent in isolated hexamers or disordered network patches (cf. Supplementary Fig. 8). The second indication is given by the emergence of a peak with onset at 464 mV, followed by three further peaks when sweeping up to 1000 mV (respectively indicated by a red arrow and black bars in Fig. 3a), which energy positions correspond neither to confined SS electrons nor to single Ext-TEB units, being featureless in this bias range[32]. The spectrum taken on the alkynyl-Ag atoms shows merely a steadily increasing intensity (Fig. 3a). Figure 3b displays d$I$/d$V$ spectra taken at the same positions with a larger bias range. Again, their shapes do not vary at equivalent sites. The averaged spectrum on Ext-TEB units peaks at ~2200 mV, whereas for the Ag-bridge sites a single broad peak dominates at ~2050 mV, markedly downshifted when compared to the characteristic signature at 2.9 V of an individual Ag adatom on Ag(111)[39].

Real-space electronic states associated with SS confinement and intrinsic Ag-GDY band formation appear clearly in bias-dependent d$I$/d$V$ maps (Fig. 3d) with energy positions indicated in Fig. 3a,b. At low biases, the imaged local density of states (LDOS) within the nanopore reflects typical intensity patterns of confined SS electrons (n=1 and 2; Fig. 3d-1,2). For a bias $V_B$ of 464 mV (Figs. 3d-3), localized LDOS intensity clearly appears on the triangular-shaped Ext-TEB backbone, whence it is associated with the conduction band minimum (CBM; cf. Fig. 3a). With further increasing bias the LDOS extends to the Ag-bridge sites, until a uniform honeycomb grid evolves at 677 mV, indicating delocalized electronic character (Figs. 3d-4,5 and Supplementary Fig. 9). From 756 mV to 955 mV, the LDOS concentrates at the alkynyl-Ag atoms (Fig. 3d-6,7). Most intriguing is the bias range from 1053 mV to 1355 mV, where the electron density at the molecular edge becomes prominent and merges via the alkynyl-Ag sites to form a marked Kagome grid at 1355 mV (Fig. 3d-8-10 and Supplementary Fig. 10). Examining the even higher bias range from 1.5 V to 2.2 V, the electronic Kagome grid becomes gradually filled until a nodal plane develops at the Ag-bridge (Fig. 3d-11-14).

The continuous evolution of LDOS pattern following the network topology manifests band characteristics in the Ag-GDY sheet. To search for the VBs (Fig. 1c), STS measurements also explored the bias range below the Fermi level, however, no significant features were observed (see Supplementary Fig. 11), similarly to other reported covalent structures adsorbed on metal substrates[33,40].

**ARPES observation of valence-band formation.** To probe the electronic properties of Ag-GDY in the occupied states regime, complementary ARPES measurements were carried out for networks grown on Ag(111) providing a sharp LEED pattern (cf. Fig. 2b and Supplementary Fig. 12). As reference the band structure $E(k_y)$ of a clean substrate measured along the $\overline{\Gamma K}$ direction of the Ag(111) surface Brillouin zone (BZ) with a photon energy of 62 eV is depicted in Fig. 4a. The well-known Ag SS surrounded by the projected bulk bands as well as the highly-dispersive $sp$-bands prevail. Since the silver $d$-bands' onset energy is 3.5 eV below the Fermi level (cf. Supplementary Fig. 13), a wide energy window for screening of Ag-GDY bands exists. Next, the occupied frontier orbitals of the intact Ext-TEB molecules were examined using an organic multilayer sample, whereby the nondispersive HOMO and HOMO-1 levels at $E_b \approx 2.8$ and 3.2 eV can be clearly discerned with maximal intensity at ∼1.5 Å$^{-1}$, i.e., no delocalized states exist for the supramolecular assembly[41,42] (Figs. 4b and Supplementary Fig. 14).

In Figure 4c, the band structure for the Ag-GDY fully covering the surface is shown. The previously distinct molecular orbitals of individual Ext-TEB (Fig. 4b) transform into less prominent intensity features smeared out along the $\overline{\Gamma K}$ direction in an energy window between $E_b \approx 2$ and 3 eV (cf. also Supplementary Fig. 15c). The VB formation is clearly recognized in Fig. 4d when comparing the energy distribution curves (EDCs) of pristine molecules and Ag-GDY extracted in the vicinity of $k_y$=1.35 Å$^{-1}$ in Fig. 4a-c. The HOMO (H) and HOMO-1 (H-1) peaks (grey curve in Fig. 4d) are well separated above the onset of the *d*-bands (purple area at $E_b > 3.8$ eV), in agreement with the DFT calculation (Fig. 1a). By contrast, for the adsorbed Ag-GDY sheet, a wider double-peak feature with an onset shifted closer to the Fermi level at $E_b \approx 2.0$ eV appears, followed by a much broader increase merged into the silver *d*-bands ($E_b \approx 3.8$ eV). The first set of VBs (denoted by $VB_1'$) has a relatively wide bandwidth of ~0.8 eV with clear intensity variations, indicating a fine structure due to multiple band contributions (cf. also Supplementary Fig. 15c). The higher intensity appearing at increased binding energies is attributed to the next set of bands denoted by $VB_2'$. Obviously, the valence band maximum (VBM) of the adsorbed system shifts below the Fermi level compared to the freestanding layer (Fig. 1c), for reasons to be clarified below. Moreover, in Fig. 4e ARPES data for a photon energy of 100eV are reproduced such that the projection of the bulk bands around the $\overline{\Gamma}$ point is strongly attenuated, leading to a better recognition of the VBs. It is evident that $VB_1'$ and $VB_2'$ do not change their position with photon energy, whereas the intensity and the *E*(*k*) distribution of the highly dispersive silver *sp*-bands are altered (cf. also Supplementary Fig. 16)[43], substantiating the conclusion that these VBs originate from the Ag-GDY sheet[44].

**DFT modelling of adsorbed Ag-GDY network.** To develop a better understanding of the experimental results, we carried out systematic DFT calculations of Ag-GDY sheet on a thick Ag(111) slab (cf. Methods). Fig. 5a displays the projected density of states (PDOS) of the system. In the unoccupied regime, the energy positions of CBM, the PDOS peak at 1 eV and the double-peak feature at 2 eV agree well with the STS observations (cf. Figs. 3a,b and 5a). In the occupied regime, the VBM sets in at -1.2 eV, followed by two main peaks at -1.7 eV and -2.7 eV (Fig. 5a), which can be nicely correlated with the $VB_1'$ and $VB_2'$ in the ARPES data (cf. Fig. 4d). Note that the combined evidence from STS and ARPES data defines a gap size of 2.5 eV. It is well known that DFT calculations usually underestimate gap sizes, whence the obtained DFT value of $E_g \approx 1.7$ eV is rather satisfactory. Fig. 5b displays a series of simulated d$I$/d$V$ maps. There is clearly a good match between experimental and simulated images in the appropriate energy ranges. Firstly, the confined Ag SS resonances ($n = 1,2$) in the nanopore are nicely reproduced due to the employed thick Ag slab (Fig. 5b-1,2; cf. also Supplementary Fig. 17). Secondly, typical bias-dependent LDOS patterns of the network are captured: starting from localized density on the Ext-TEB core (Fig. 5b-3) to displaying a uniform honeycomb electronic pathway (Fig. 5b-4), followed by merging of Ag-bridge (Fig. 5b-5) and molecular edge state (Fig. 5b-6) into a real-space electronic Kagome grid (Fig. 5b-7). Overall, the DFT calculations of the full system are in good agreement with the experimental observations.

Compared to the freestanding network, the VBs of the adsorbed layer on the silver surface notably shift below the Fermi level, indicating that the open shell of the unit cell is filled. An estimate obtained with the Bader charge analysis identified a total charge transfer of $\approx 3$ electrons from the substrate as well as the bridging silver adatom to compensate six alkynyl radicals in the unit cell. Moreover, the absence of well-defined Kagome-type $E(k)$ relations in the ARPES data can be ascribed to a mild electronic hybridization between the adlayer and the silver surface (cf. Supplementary Fig. 18-20) as well as to a small BZ of the network. Nevertheless, at CBM ($V_B = 464$ mV) the LDOS intensity localized on the molecules captured in the experimental as well as simulated d$I$/d$V$ maps (Figs. 3d-c and 5b-3) is compatible with the nontrivial-flat-band character, namely, quenched electron kinetic energy due to destructive interference of Bloch waves[14].

Finally, in order to determine whether the intrinsic Ag-GDY electronic bands would be retained on an insulator, a sheet supported on a *h*-BN monolayer was modelled (Fig. 6a). Importantly, the leading VBs remain identical as those of the freestanding layer, and $CB_1$ and $CB_2$ slightly shift down toward the Fermi level (Fig. 6c). Notably, a finite (~ 37 meV) gap opens at the Γ point, which gives rise to an isolated flat band at the bottom of $CB_1$ (cf. red solid lines in Fig. 6c). Also, gap-opening at Dirac point can be observed for both $CB_1$ (~ 30 meV) and $CB_2$ (~ 37 meV). Note that the phenyl rings of individual Ext-TEB molecule in the unit cell locate on either nitrogen or boron atoms (Fig. 6a). Therefore, the adsorption-induced geometrical symmetry breaking provides another effective way to open band gaps[14,16].

**Discussion**

The realization of high-quality crystalline Ag-GDY sheets permitted an in-depth electronic structure characterization at both the atomic and mesoscopic scales. Band formation with a semiconducting gap of ≈2.5 eV was rigorously determined in this π-conjugated GDY analogue. Moreover, we reveal that Ag-GDY hosts unconventional flat, Dirac and Kagome bands at both conduction and valance band edges, which originate from combined MO and lattice symmetries. Although mild hybridization between the network and metal surface interferes, we show the possibility to restore the intrinsic band structure via electronic decoupling. Based on our results, replacing the molecular units and different metallation schemes can be envisioned. Therefore, the proper combination of MO characteristics and lattice symmetry can give access to bespoke novel electronic structures, unfolding exciting opportunities for tailored 2D material developments amenable to device-integrated applications.

**Methods**

**Sample preparation**. The Ag-GDY networks were prepared under ultrahigh vacuum conditions. Ag(111) single crystals and Ag(111)/mica substrates were used for sample growth. 1,3,5-tris(4-ethynylphenyl)benzene (Ext-TEB) precursors were evaporated from $Al_2O_3$ or quartz crucibles onto clean Ag(111) surfaces held at 200 K ~ 300 K, followed by an alkyne deprotonation procedure via introducing $O_2$ gas (≈ 6000 L) into the preparation chamber by backfilling via a leak valve. The sample was mildly annealed ($T_{ann}$ ≈ 400 K) to facilitate the organometallic network formation.

**STM/STS characterization.** A commercial Joule-Thomson STM (JT-STM, SPECS) and a homemade LT-STM were used for data acquisition. Data was recorded at an equilibrium temperature of ~5 K. Point STS spectroscopy were measured via a lock-in amplifier with a bias modulation of 20 mV at 676 Hz. d$I$/d$V$ maps were recorded at fixed bias voltage in a grid with 128 × 128 pixels, and the feedback was switched off during d$I$/d$V$ signal acquisition at each pixel. All the d$I$/d$V$ maps were normalized via applying (d$I$/d$V$)/($I$/$V$), which leads a better signal to noise ratio without altering the LDOS features.

**ARPES measurements.** The ARPES experiments presented in this work were carried out at the Photo-Emission and Atomic Resolution Laboratory (PEARL) of the Swiss Light Source (SLS). The PEARL beamline delivers tunable soft X-ray photons in the energy range from 60 to 2000 eV with a resolving power up to E/ΔE=7000. In the present work we used photon energies of 62 eV and 100 eV to maximize the surface sensitivity. Samples are mounted on a high-precision manipulator with three translation and three rotation axes. Photoelectrons are detected by a Scienta EW4000 wide acceptance angle analyser with a two-dimensional multi-channel plate detector where one axis corresponds to the kinetic energy of the electron and the other axis to the emission angle. The entrance lens stack of the analyser is at a fixed angle of 60° with respect to the incoming synchrotron light. The X-ray beam, the polarization vector of photons, and the axis of the analyser lens are oriented horizontally, while the entrance slit of the electron analyser is oriented vertically. Samples were measured either at RT or at 60 K.

**DFT calculations.** Total energy calculations were performed using DFT within the Kohn–Sham formalism[45] using the Quantum ESPRESSO code[46]. We used the rB86-vdW-DF2 approximation in the term of the exchange-correlation functional[47], and only the $\overline{\Gamma}$ point in the integration over the first Brillouin zone due to the very large super-cell. Projector augmented wave[48] data sets were applied to remove the core electron from the explicit calculation. STM simulations were performed with the Tersoff–Hamann model[49] approximating an *s*-wave tip.

We performed the calculations of the supported network in a hexagonal super-cell, unlike in Ref [31], where we used a rectangular, elongated cell. In order to reduce the stress in the overlayer we slightly adjusted the lateral lattice constant of the substrate to 4.18794 Å so that the average distance between the bridging Ag ad-atoms was the same as in the large system. During the relaxation and analysis of the electronic structure we used four layers of the substrate and a passivating layer of hydrogen at the bottom of the slab, and each layer of the substrate and the passivating layer consisted of 147 atoms, yielding a lateral lattice constant of 35.904 Å. In the calculations of the band structure, we could afford to include only two top-most layers of substrate layers. The cut-off energies for wave functions and the electron density were 35 and 500 Ry, respectively.

In the calculation of the Ag-GDY network on the *h*-BN, 196 unit cells of latter were included with the same Ag-GDY network parameters as those for the silver substrate, whereby all the atoms were allowed to relax in any direction.

**Acknowledgements**
This work is financially supported by the National Natural Science Foundation of China (12174431 and 11825405) and the Strategic Priority Research Program of the Chinese Academy of Sciences (XDB30000000). J.V.B. acknowledges support from ERC (Advanced Grant MolArt), and the Munich Quantum Center, and the DFG Excellence Cluster e-conversion. F.H., J.K and F.A. acknowledge funding by the German Research Foundation (DFG) through the TUM International Graduate School of Science and Engineering (IGSSE, GSC81). We gratefully acknowledge the Paul Scherrer Institut (Villigen, Switzerland) for the provision of synchrotron radiation beamtime at beamline PEARL of the Swiss Light Source (SLS). We also thank Willi Auwärter for helpful discussions.

**Author contributions**
Y.-Q.Z., I.P.-Z. and J.V.B. conceived the experiments. I.P.-Z. led the ARPES measurements and analysed the data with contribution from Y.-Q.Z., F.H., J.K. F.A. and M.M.. W.H., Y.L., Y.-Q.Z., L.C., and K.W. performed the STM/STS experiments and analysed the data. A.P.S. carried out the DFT calculations. Z.M.A.E.-F. carried out confined-surface-states analysis. E.A. performed tight-binding modelling. S.K. and M.R. developed the synthesis of the molecules used. Y.-Q.Z., I.P.-Z., A.P.S. and J.V.B. co-wrote the paper with contribution from all authors.
**Competing interests**
The authors declare no competing interests.

**Additional information**
Supplementary information is available for this paper.

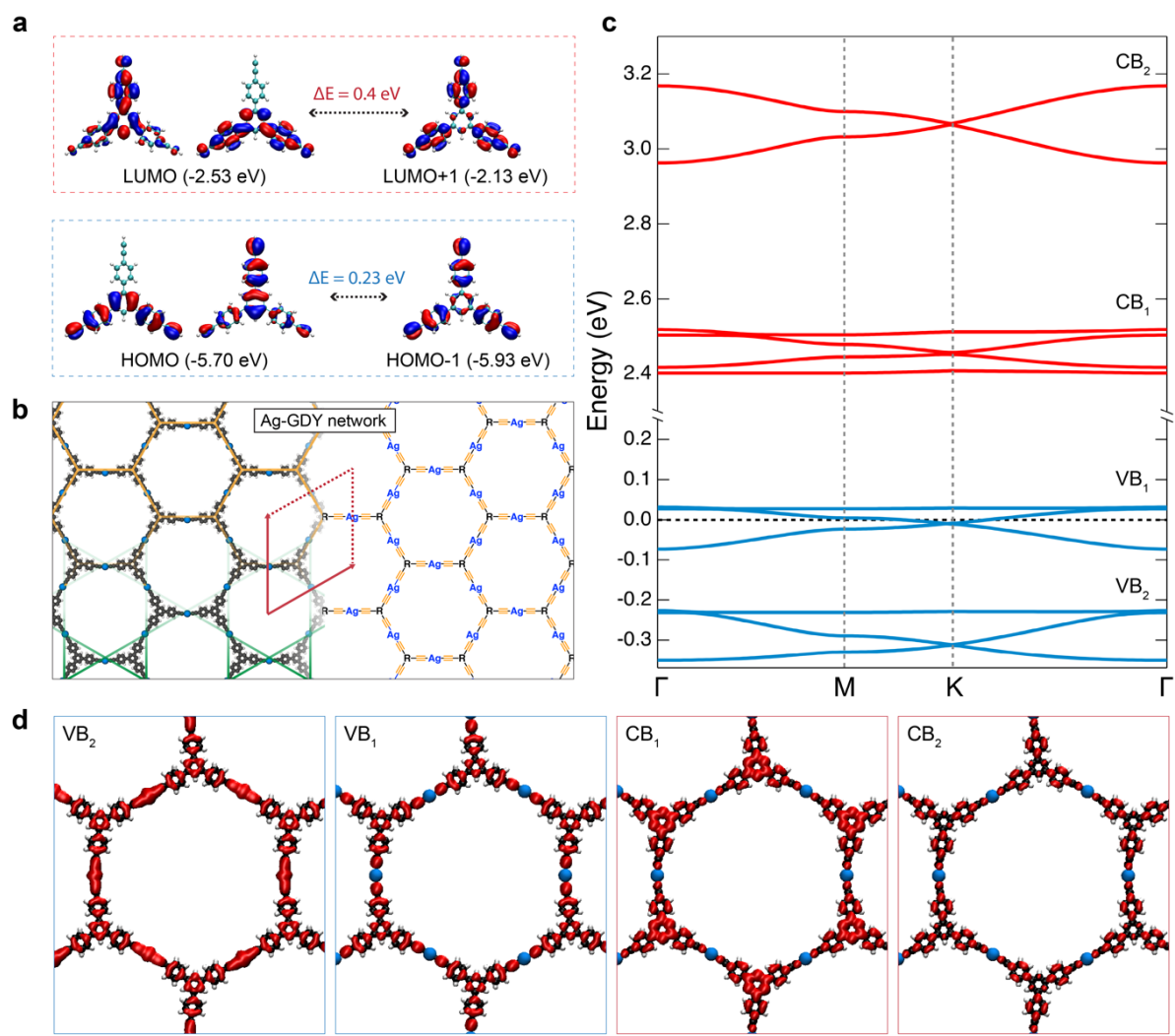

**Fig. 1 | Doubly degenerate precursor LUMO and HOMO account for flat, Dirac and Kagome bands in the Ag-GDY sheet. a**, DFT calculated frontier molecular orbitals of isolated Ext-TEB molecule. **b**, Model and chemical scheme of freestanding Ag-GDY sheet with rhombic unit cell. Honeycomb and Kagome lattices occupied by molecules and Ag atoms are highlighted, respectively. **c**, Band structure around the Fermi level and above the band gap of the freestanding network. Red and blue colours are used to distinguish CBs and VBs. **d**, Integrated network orbital density of the first and second groups of VBs and CBs, respectively.

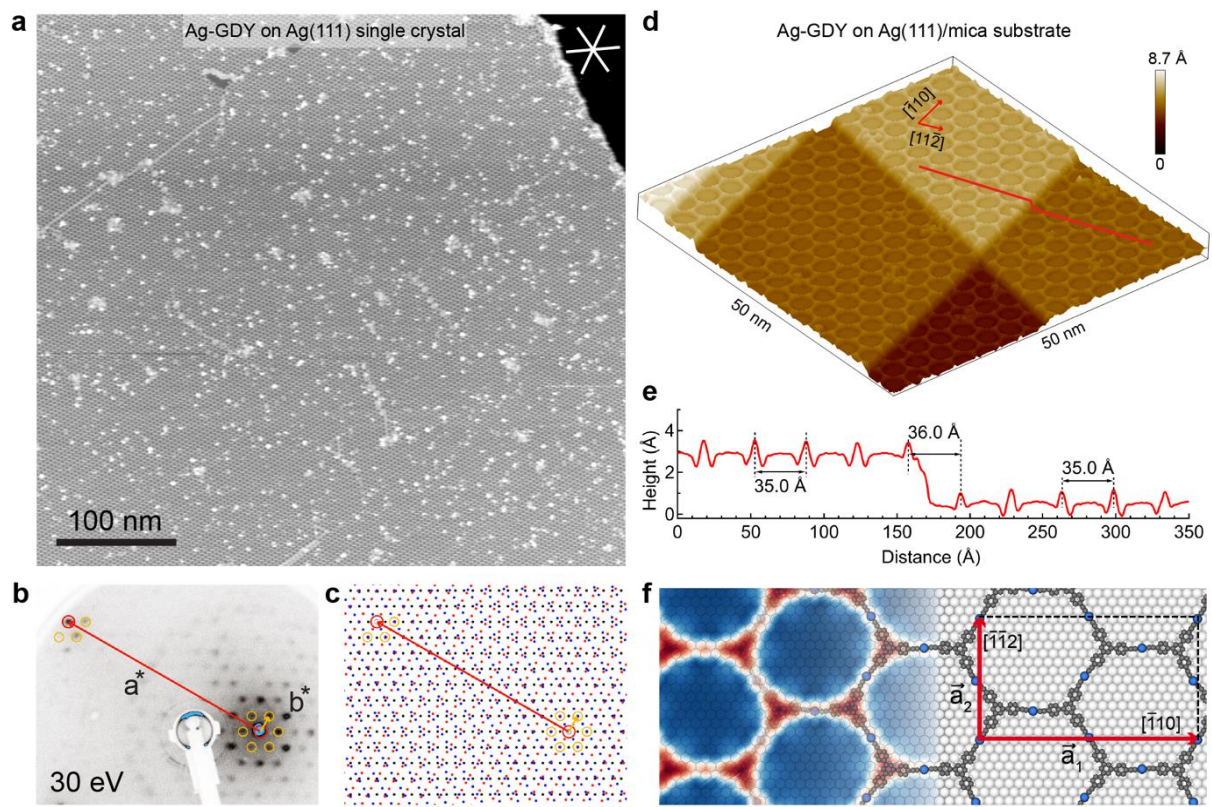

**Fig. 2 | Overview STM images of Ag-GDY network grown on Ag(111) facets. a**, STM image of a continuous 500 × 500 nm² atomically thin Ag-GDY network grown on an Ag(111) single crystal. $I_t$ = 100 pA, $U_b$ = -1.0 V. **b**, LEED pattern of Ag-GDY/Ag(111) for $T_{sub}$ = 90 K and $E_{electron}$ = 30 eV. **c**, Modelling of the LEED pattern. **d**, 3D rendering of an STM image showing the continuity of the network crossing straight steps on Ag(111)/mica. $I_t$ = 50 pA, $U_b$ = 0.1 V. **e**, STM contour of line marked in (**d**). **f**, Structure model of Ag-GDY network with rectangular unit cell on Ag(111) facet.

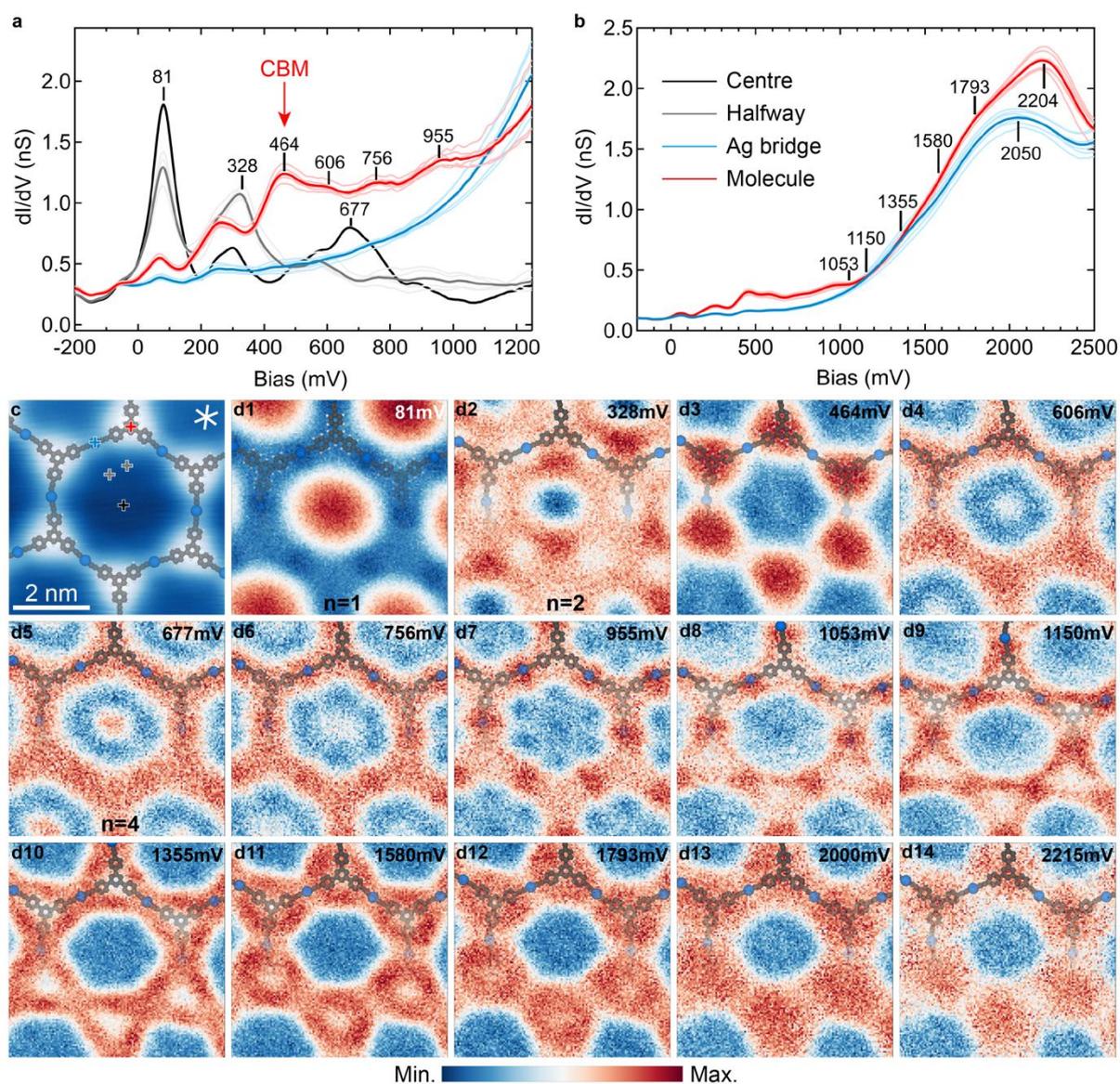

**Fig. 3 | STS characterization of the elementary constituent in a Ag-GDY/Ag(111) sheet. a-b,** Site-specific d$I$/d$V$ spectra taken in the bias range from -0.2 to 1.2 V and extended bias range from -0.3 to 2.5 V, respectively. Set point: $I_t$ = 50 pA, $U_b$ = -0.2 V and -0.3 V, $U_m$ = 20 mV. **c,** Corresponding STS survey area in the network, superimposed with a DFT model with rectangular unit cell. $I_t$ = 50 pA, $U_b$ = 0.96 V. **d,** Bias-dependent d$I$/d$V$ mapping of the regular six-membered ring shown in (**c**). Set point: $I_t$ = 50 pA, $U_m$ = 20 mV.

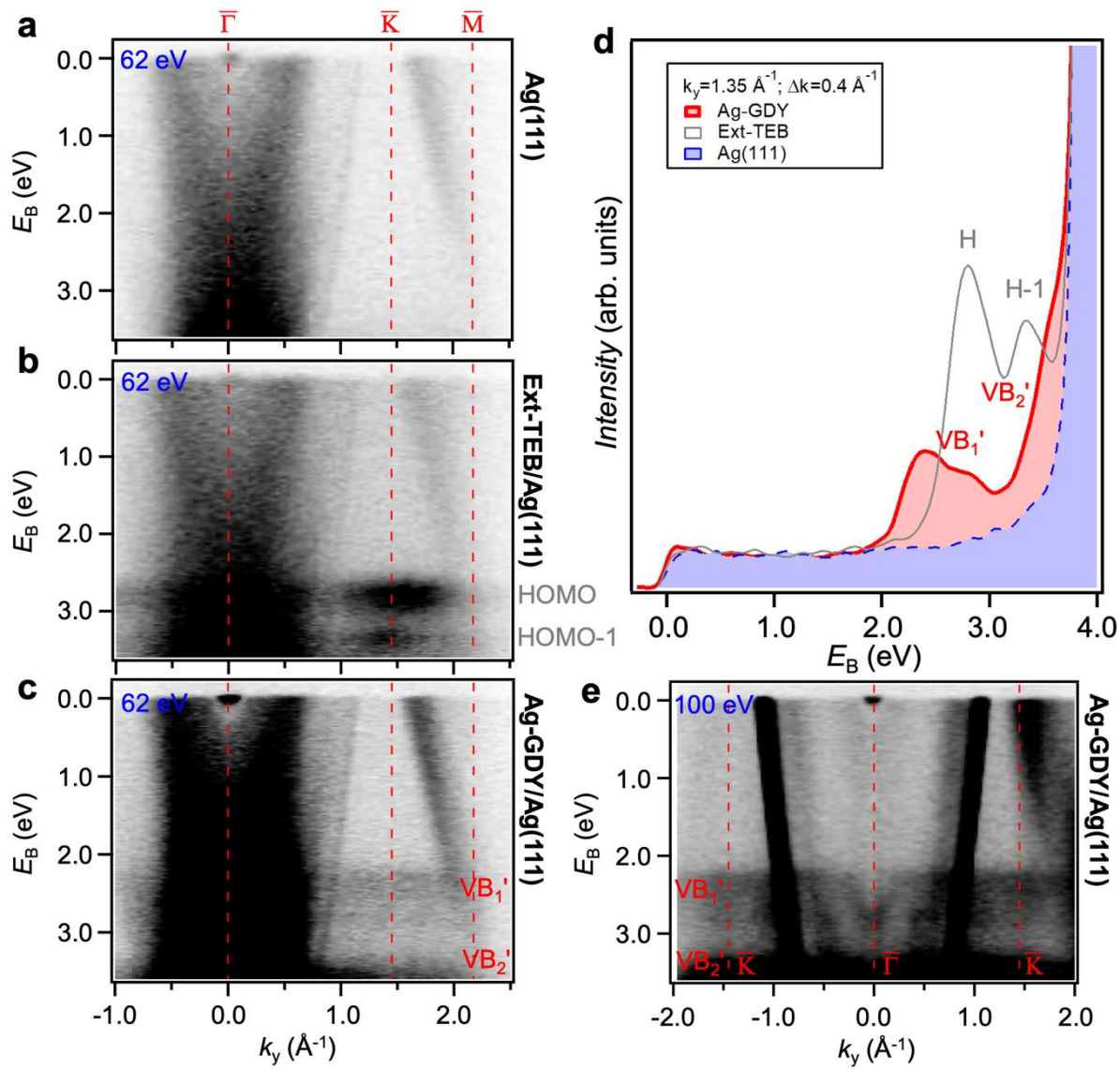

**Fig. 4 | Evolution of ARPES band structure upon Ag-GDY sheet formation. a-c**, Band structure ($E_B$ vs. $k_y$) of pristine Ag(111), Ext-TEB organic multilayer and Ag-GDY network measured along the $[1\bar{1}0]$ substrate direction (i.e. along $\overline{\Gamma K}$) with a photon energy of 62 eV. **d**, EDCs at $k_y$=1.35 Å$^{-1}$ evidence discrete occupied frontier orbitals for pure Ext-TEB molecules and their transformation to a valence band ($VB'_1$ with a ~0.8 eV bandwidth) upon Ag-GDY network formation. The additional onset close to the Ag *d*-bands corresponds to the $VB'_2$. **e**, Same band structure as in (**c**) measured with 100 eV photon energy: While Ag-*sp* bands shift with photon energy, the Ag-GDY bands remain unchanged in accordance with their 2D nature. ARPES experimental datasets in (**a**) and (**b**) were obtained at RT, (**c**) and (**e**) at T ≈ 60 K.

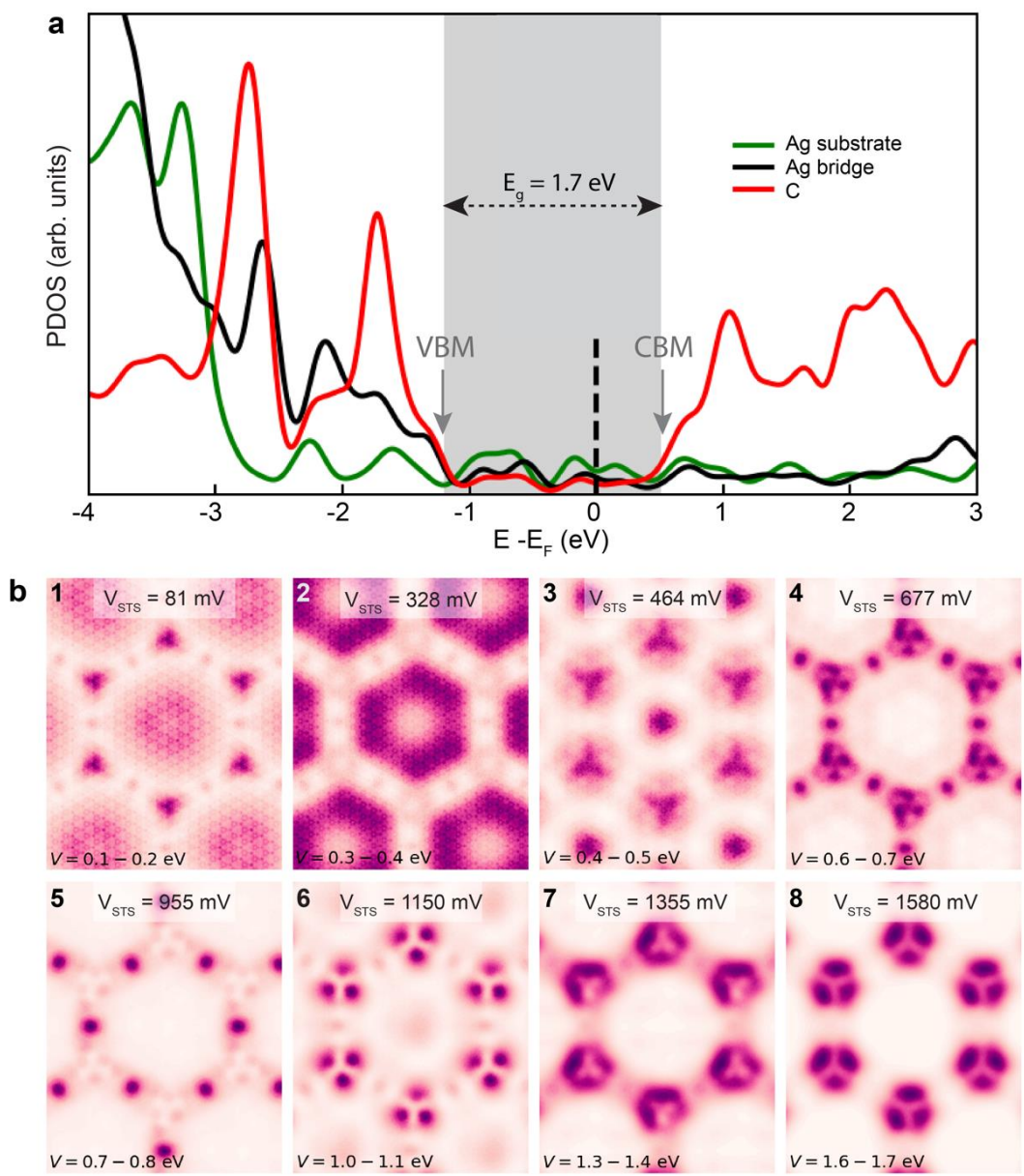

**Fig. 5 | DFT modelization of Ag-GDY/Ag(111) electronic features. a**, Projected DOS of the Ag-GDY network adsorbed on four-layer Ag(111) slab. Contributions from carbon, alkynyl-Ag atoms and substrate are distinguished by different colours. **b**, Simulated *dI/dV* maps with characteristic LDOS patterns corresponding to the experimental bias values, indicated on the top of individual figures.

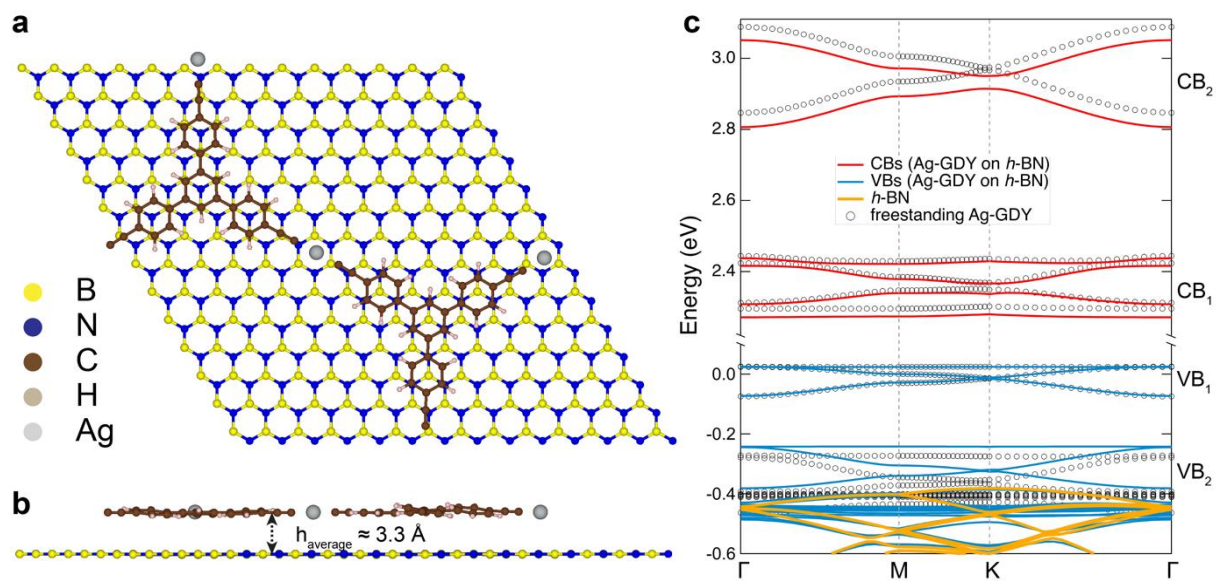

**Fig. 6 | DFT calculated band structure of Ag-GDY sheet on *h*-BN. a-b,** Top and side views of the large Ag-GDY/*h*-BN unit cell with averaged adsorption height indicated. **c**, Band structure of Ag-GDY/*h*-BN (red and blue solid lines) compared with freestanding Ag-GDY (black circles) with same geometry adopted from the combined system shown in (**a**). *h*-BN bands are highlighted in orange.

# SUPPLEMENTARY INFORMATION

# Unconventional band structure via combined molecular orbital and lattice symmetries in a surface-confined metallated graphdiyne sheet

Ignacio Piquero-Zulaica[1,†,*], Wenqi Hu[2,†], Ari Paavo Seitsonen[3,†], Felix Haag[1], Johannes Küchle[1], Francesco Allegretti[1], Yuanhao Lyu[2], Lan Chen[2], Kehui Wu[2], Zakaria M. Abd El-Fattah[4], Ethem Aktürk[5], Svetlana Klyatskaya[6], Mario Ruben[6,7], Matthias Muntwiler[8], Johannes V. Barth[1,*] and Yi-Qi Zhang[2,1*]

[1]*Physics Department E20, Technical University of Munich, D-85748 Garching, Germany*
[2]*Institute of Physics, Chinese Academy of Sciences, 100190 Beijing, China*
[3]*Département de Chemie, École Normale Supérieure, 24 rue Lhomond, F-75005 Paris, France*
[4]*Physics Department, Faculty of Science, Al-Azhar University, Nasr City, E-11884, Cairo, Egypt*
[5]*Department of Physics, Adnan Menderes University, 09100 Aydin, Turkey*
[6]*Institute of Nanotechnology, Karlsruhe Institute of Technology, 76344 Eggenstein-Leopoldshafen, Germany*
[7]*IPCMS-CNRS, Université de Strasbourg, 23 rue de Loess, 67034 Strasbourg, France*
[8]*Paul Scherrer Institute, Forschungsstrasse 111, 5232 Villigen PSI, Switzerland*

†These authors contributed equally.
*Email: ge46biq@mytum.de; jvb@tum.de; yiqi.zhang@iphy.ac.cn

**Index**

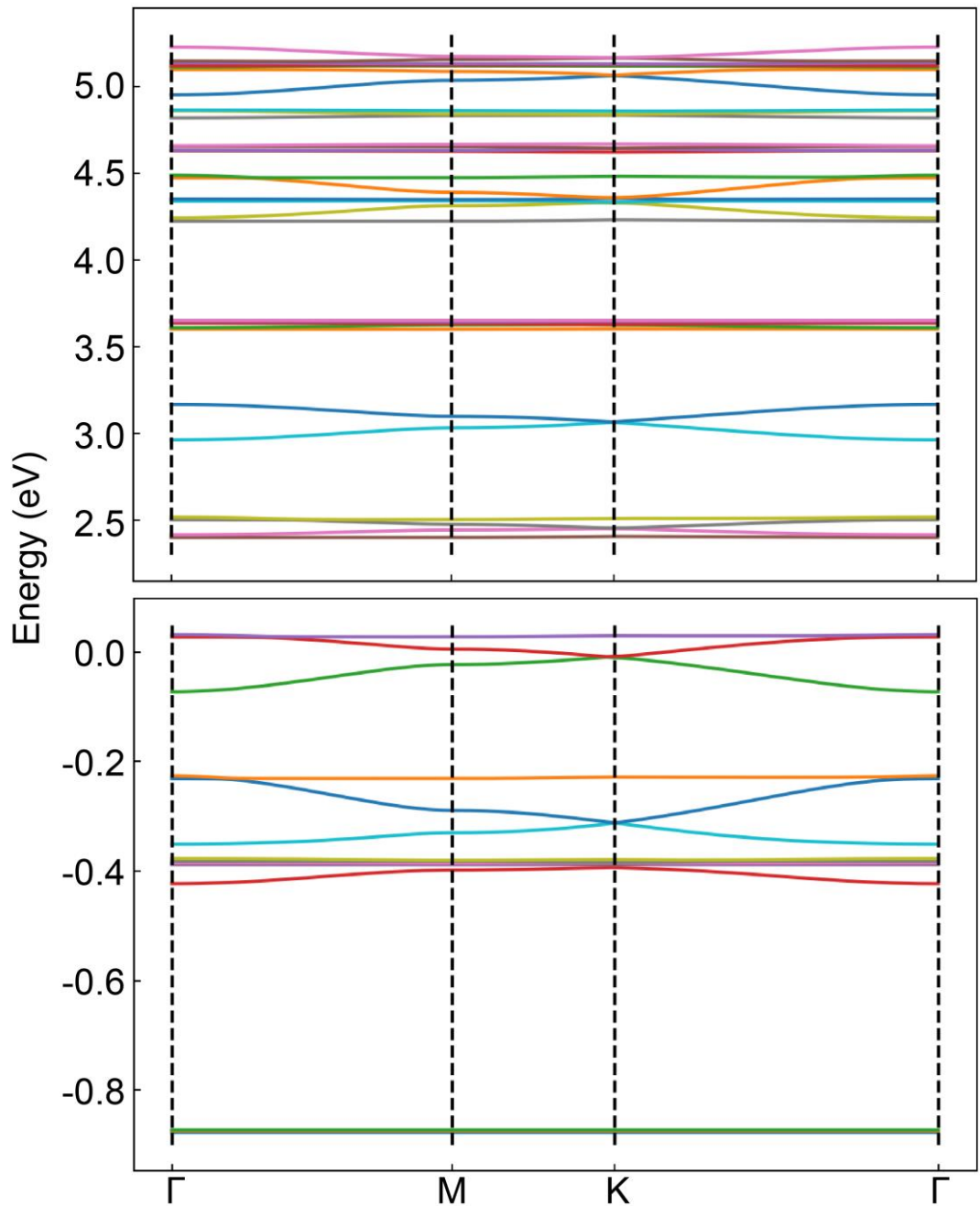

**Supplementary Figure 1 | Band structure of the freestanding Ag-GDY sheet.** The network geometry is adopted from the relaxed system with four-layer-Ag-slab, where a rhombic unit cell is employed.

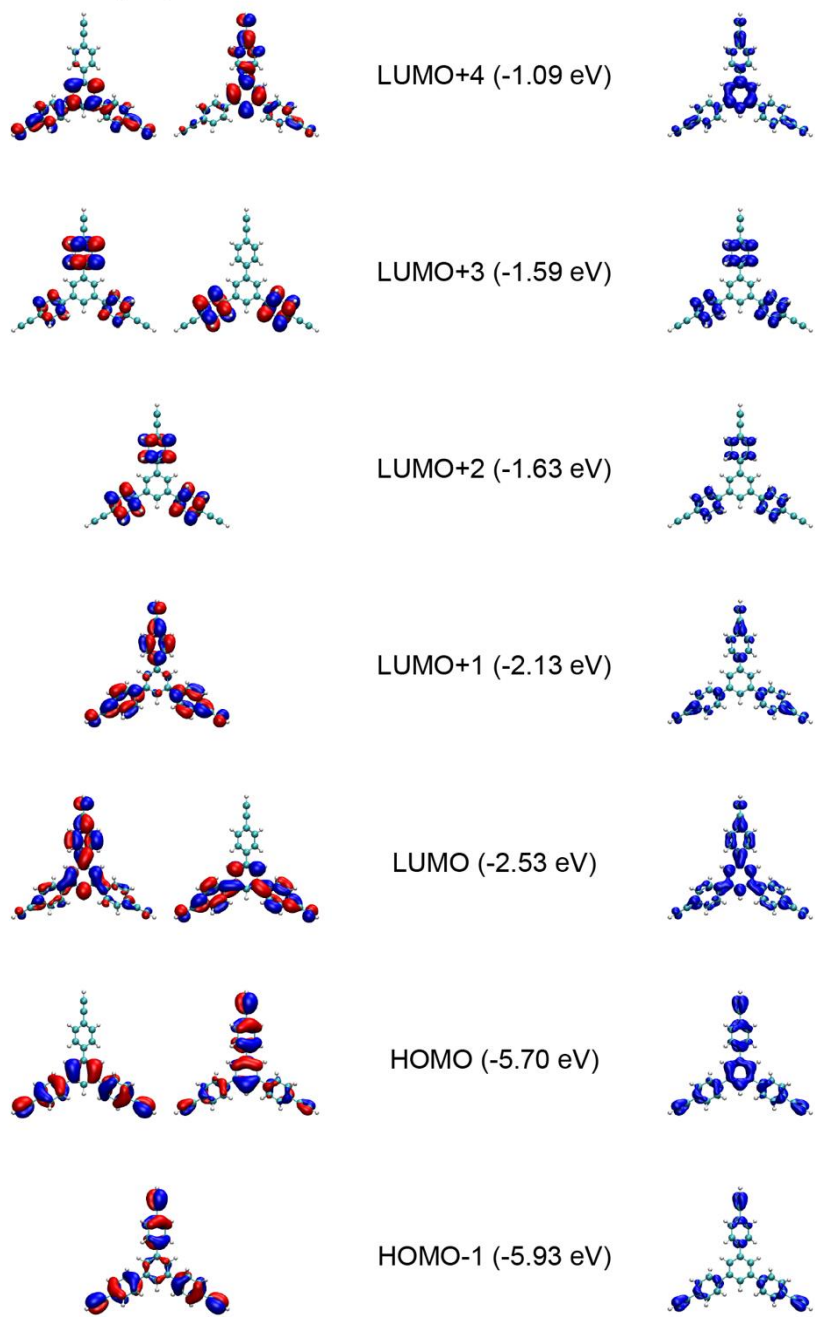

**Supplementary Figure 2 | DFT-calculated molecular orbitals of gas-phase Ext-TEB.**

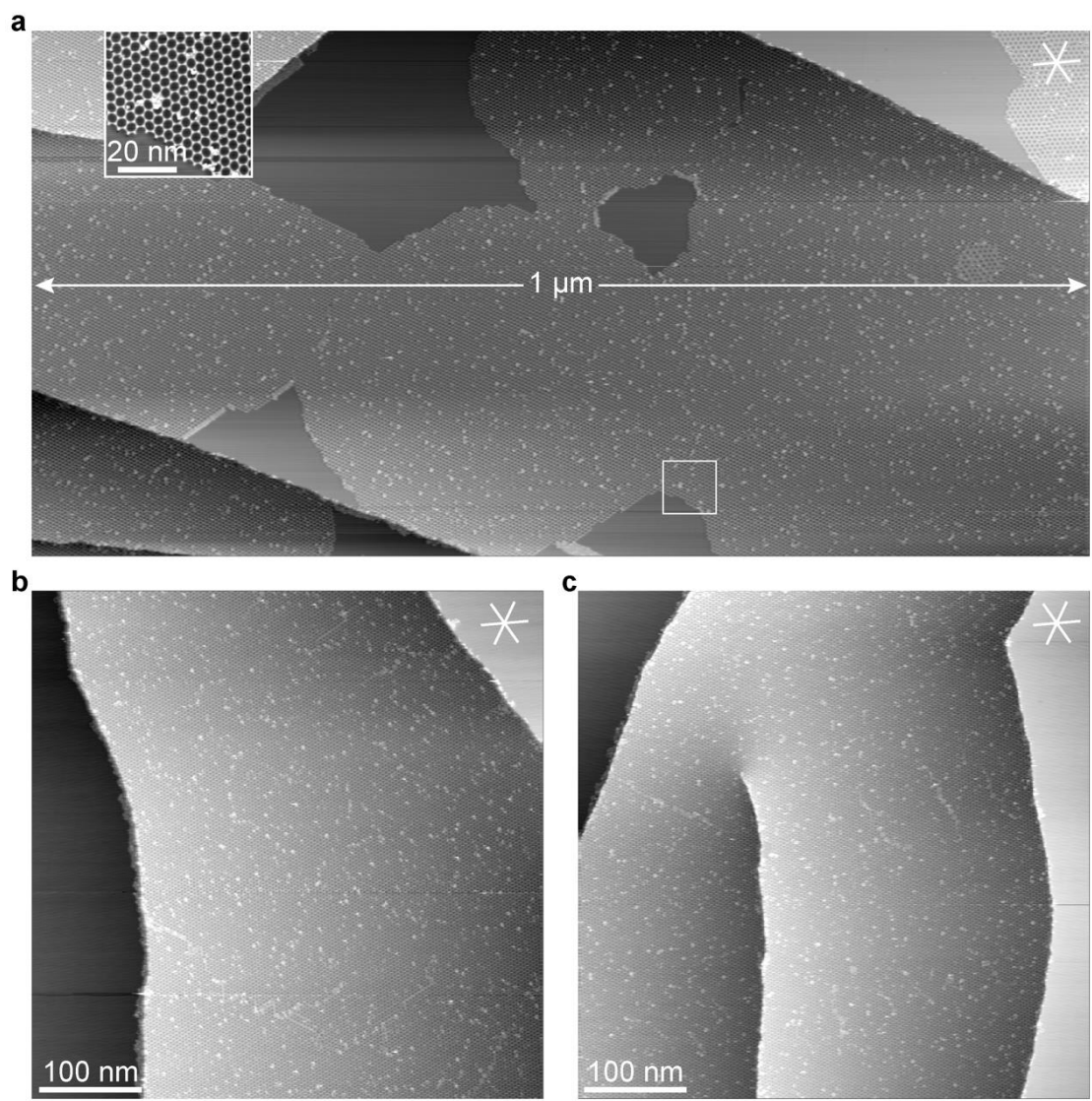

**Supplementary Figure 3 | Large-scale STM images of single-domain Ag-GDY networks grown on large terraces of the Ag(111) single-crystal surface.** Scanning parameters: (a) $I_t$ = 100 pA, $U_b$ = - 1 V. Inset: $I_t$ = 100 pA, $U_b$ = - 100 mV. (b) $I_t$ = 100 pA, $U_b$ = - 1 V. (c) $I_t$ = 100 pA, $U_b$ = - 100 mV. The bright protrusions in the networks correspond to impurities as well as trapped molecules[1].

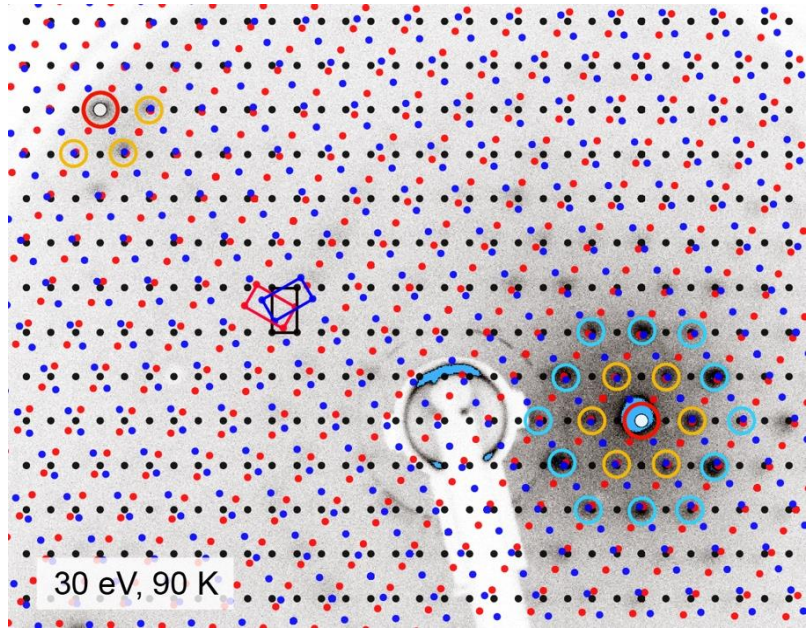

**Supplementary Figure 4 | LEED pattern modelling.** LEED pattern of the Ag-GDY sheet prepared on the Ag(111) single-crystal surface. $T_{meas}$ = 90 K, $E_{electron}$ = 30 eV. The overlayer is simulated LEED pattern generated using the *LEEDpat* software. Three sets of equivalent lattices are distinguished using red, blue and black dots, respectively. Red circles mark the diffraction spots of Ag(111). Yellow and blue circles highlight the first and the second order diffraction spots of the Ag-GDY network, respectively.

Simulated LEED pattern was generated via the *LEEDpat* software (http://www.fhi-berlin.mpg.de/KHsoftware/LEEDpat/) using the experimentally determined rectangular-unit-cell parameters[1]. Three equivalent domains with a rotation angle of 120° are taken into account. Ag(111) diffraction spots (red circles) are used for calibrating the simulated LEED pattern to be compared with the experimental LEED photo. It is clear to see that the first (yellow circles) and second (light blue circles) order diffraction spots correspond to the reciprocal spots where three sublattices present the maximum overlap.

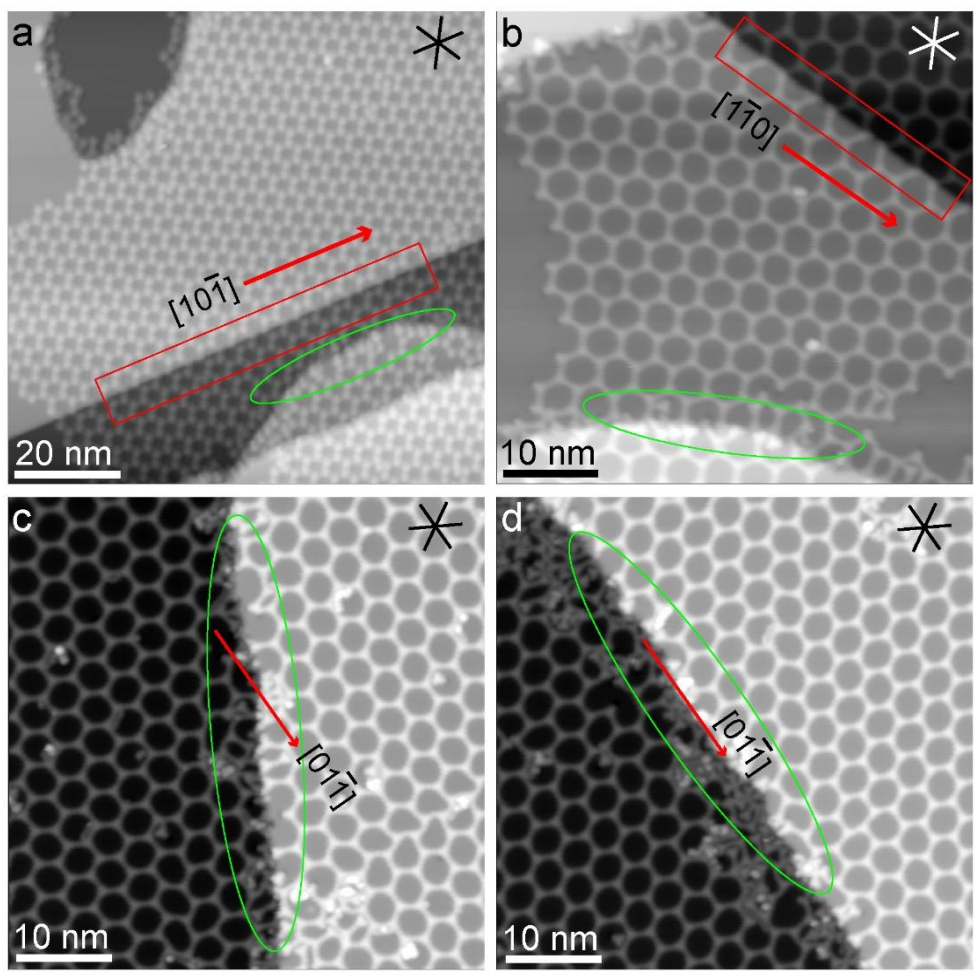

**Supplementary Figure 5 | The continuity of the Ag-GDY network at the step edges.**
(a,b) STM images of Ag-GDY network growing continuously across a monoatomic-height step along the high symmetry direction on the Ag(111)/mica substrate. (c,d) STM images of Ag-GDY networks discontinued at the monoatomic-height step edges on the Ag(111) single-crystal surface. Solid red rectangles highlight atomically straight step edges. Green ellipses highlight irregular step edges. High-symmetry directions of all the Ag(111) surfaces are indicated. Scanning parameters: (a) $I_t$ = 30 pA, $U_b$ = 1V; (b) $I_t$ = 50 pA, $U_b$ = 0.5 V; (c, d) $I_t$ = 100 pA, $U_b$ = - 1V.

On the Ag(111)/mica substrate, there exhibits atomically-sharp step edges along high symmetry directions, and we found that Ag-GDY network can grow continuously across such a monoatomic-height step (cf. also Fig. 2d in the main text). We found that the network is slightly stretched (3%) along the direction perpendicular to the step edge (Supplementary Fig. 6).

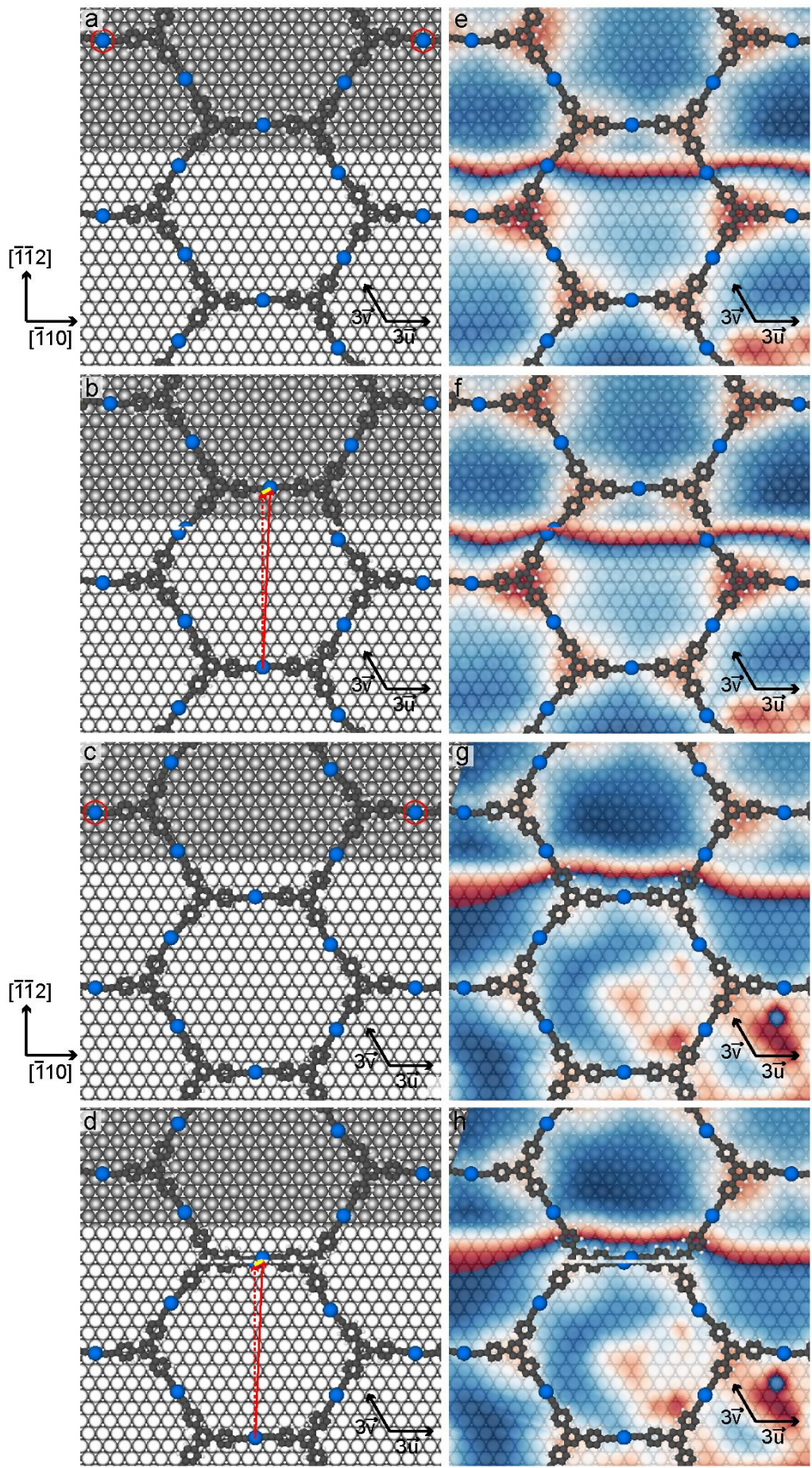

**Supplementary Figure 6 | Modelling of network distortion at step edge**. Two different adsorption geometries are analysed. Alkynyl-Ag atoms (a) or Ext-TEB molecules (c) are locating at the step edge with favourable network adsorption configuration on the upper terrace. The red circles highlight the alkynyl-Ag atom sitting on the top site. (b,d) Ideal network adsorption configurations on both terraces with a small shear present at the step edge. The red dotted and solid-line vectors show the alkynyl-Ag position before and after a small displacement, indicated by the yellow line. Four different models shown in (a-d) are superimposed with STM images and displayed in (e-h).

The adsorption registry for the Ag-GDY network growing continuously across step edge is modelled based on DFT-relaxed structure on the flat surface[1]. Supplementary Fig. 6 shows two types of step-crossing geometries, with alkynyl-Ag atom (Supplementary Fig. 6a) or Ext-TEB molecule (Supplementary Fig. 6c) locating at the step edge, respectively. If we assume that there exhibits a stretching vector $\frac{2}{3}\vec{u} + \frac{1}{3}\vec{v}$ (Supplementary Fig. 6b,d) at the step edge, all alkynyl-Ag adatoms on both terraces can be assigned to the Ag(111) hollow sites, i.e., on the favourable adsorption sites. Note that the stretching projection along the $[\bar{1}\bar{1}2]$ direction is ~0.3$a_0$ ($a_0$=2.889 Å), which agrees well the experimental data (~ 1Å) as shown in Fig. 2e in the main text.

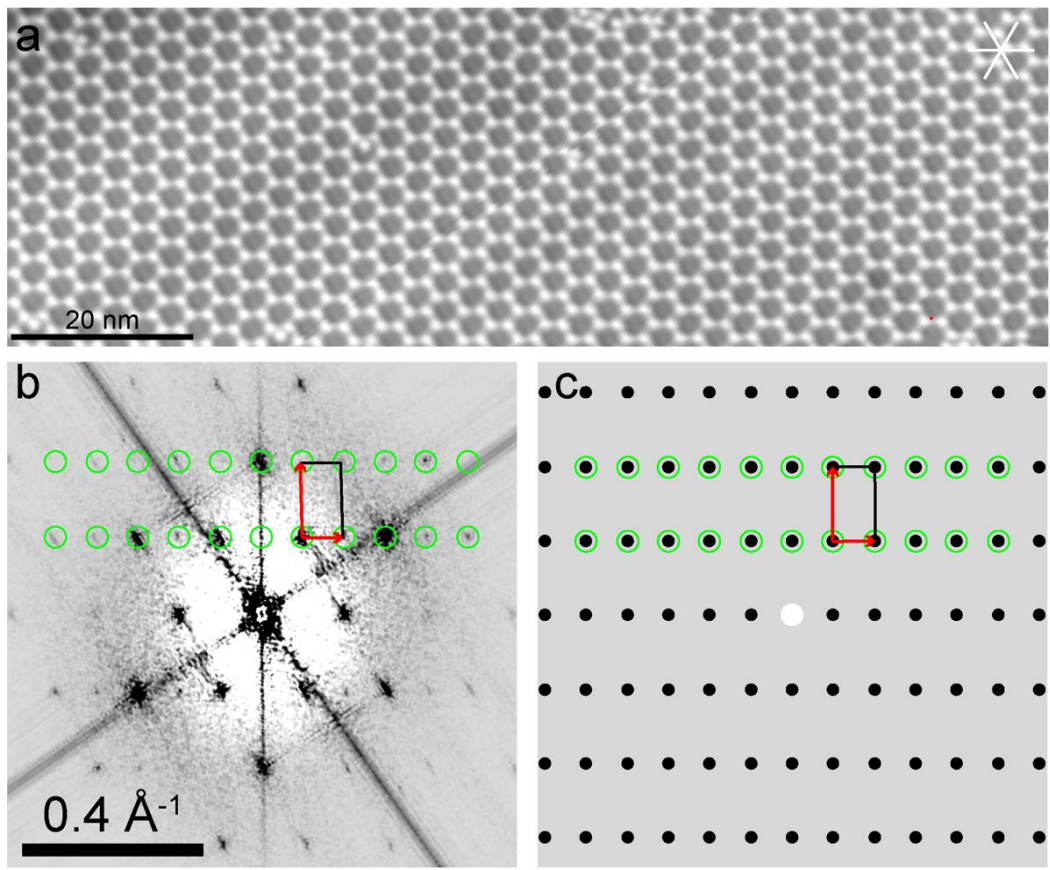

**Supplementary Figure 7 | Rectangular unit cell of Ag-GDY network grown on the Ag(111)/mica substrate**. (a) STM image of a single domain of the Ag-GDY network grown on the Ag(111)/mica substrate. High-symmetry directions of the Ag(111)/mica surface are indicated. $I_t$ = 30 pA, $U_b$ = 1.0 V. (b) The 2D fast Fourier transform (2D-FFT) of an area (120 nm × 120 nm). (c) The simulated LEED pattern of a single domain of the Ag-GDY network. Green cycles highlight rectangular reciprocal lattice of the network. The white dot denotes the reciprocal lattice spot of Ag(111).

Supplementary Figure 7a is an STM overview image of a single domain of the Ag-GDY network grown on the Ag(111)/mica substrate. We can obtain a commensurate rectangular unit cell ($a_1$ = 63.6 Å, $a_2$ = 35.0 Å), which agrees well with our previous result[1]. The periodic rectangular reciprocal lattice shown in 2D-FFT image (cf. Supplementary Fig. 7b) obtained from a large single domain corroborates the rectangular unit cell and agrees with the simulated reciprocal lattice employing the *LEEDpat* software.

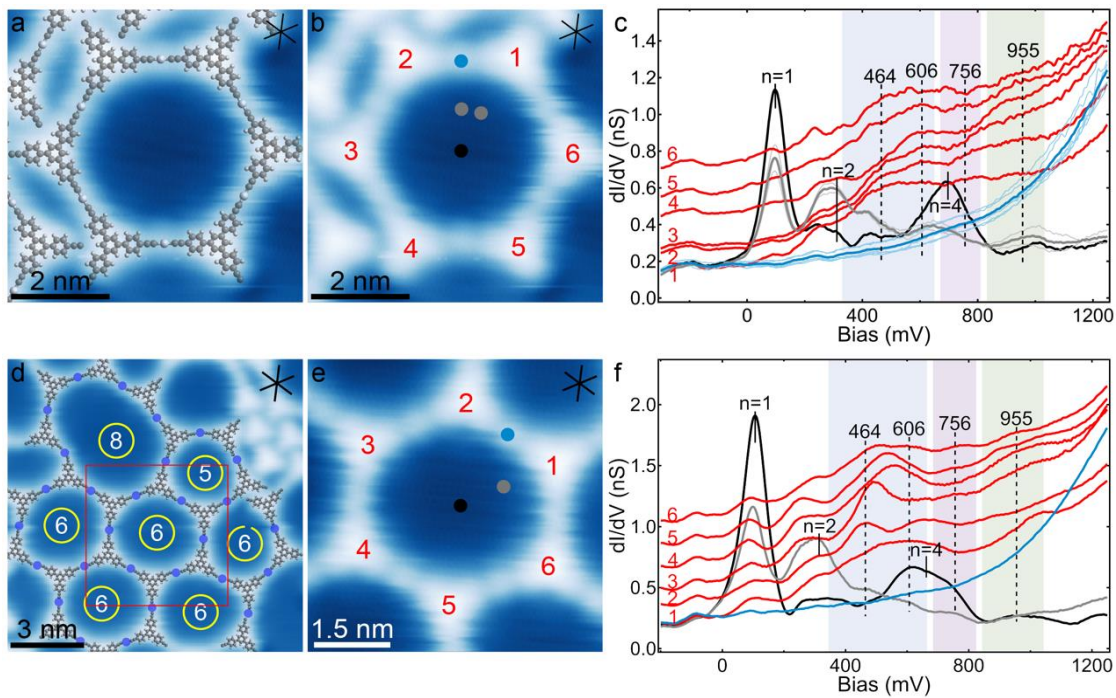

**Supplementary Figure 8 | d*I*/d*V* characterization of discrete and disordered organometallic network on Ag(111)/mica.** High-resolution STM image of a six-membered organometallic ring superimposed with the model (a) and point-spectra sites (b). (c) Site-dependent d*I*/d*V* spectra of the organometallic hexamer. Set point: $I_t$ = 50 pA, $U_b$ = - 300 mV, $U_m$ = 20 mV. (d) STM image of a disordered network. Yellow cycles and numbers highlight the surrounding numbers of the molecules. (e) Magnified STM image of (d) with six-membered ring embedded in the disordered network with point-spectra sites indicated and corresponding d*I*/d*V* spectra displayed in (f). Set point: $I_t$ = 50 pA, $U_b$ = - 200 mV, $U_m$ = 20 mV.

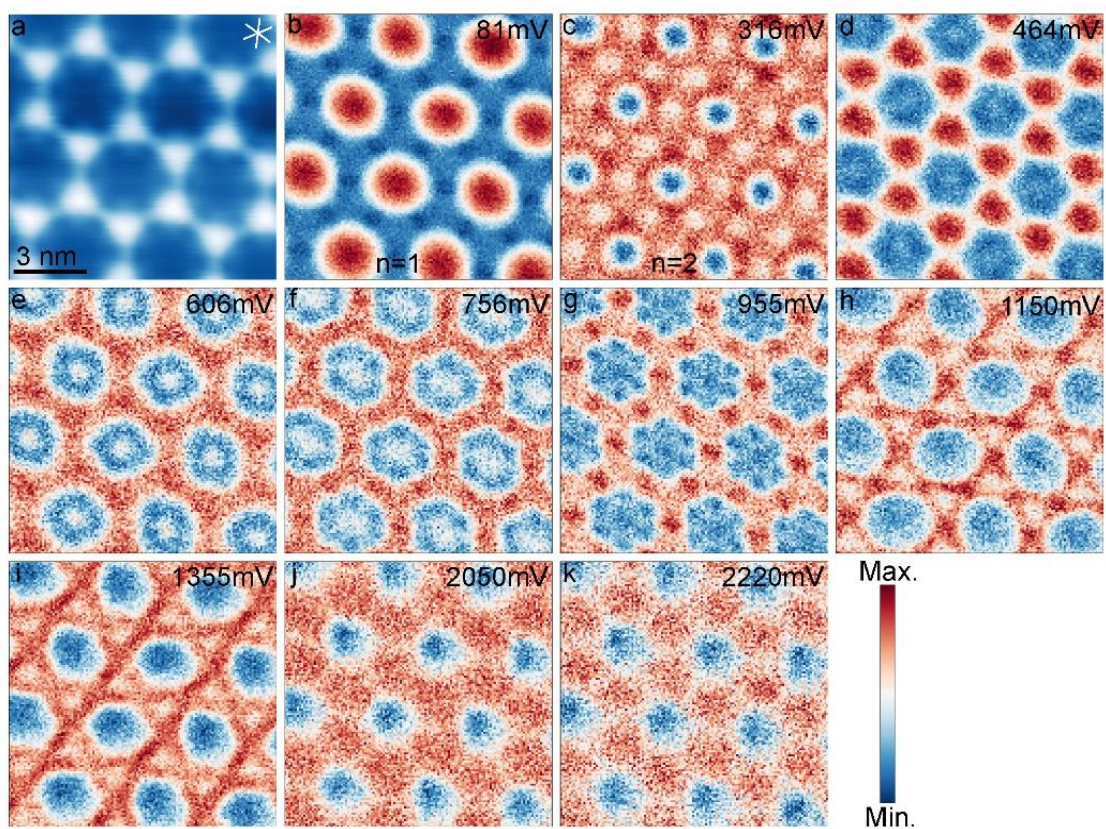

**Supplementary Figure 9 | Larger-scale d$I$/d$V$ maps of the Ag-GDY network at the unoccupied regime.** (a) Larger-scale STM image of Ag-GDY network. $I_t$ = 50 pA, $U_b$ = 606 mV. High-symmetry directions of the Ag(111)/mica surface are indicated. (b-k) Corresponding d$I$/d$V$ maps at different bias voltages. Set point: $I_t$ = 50 pA, $U_m$ = 20 mV.

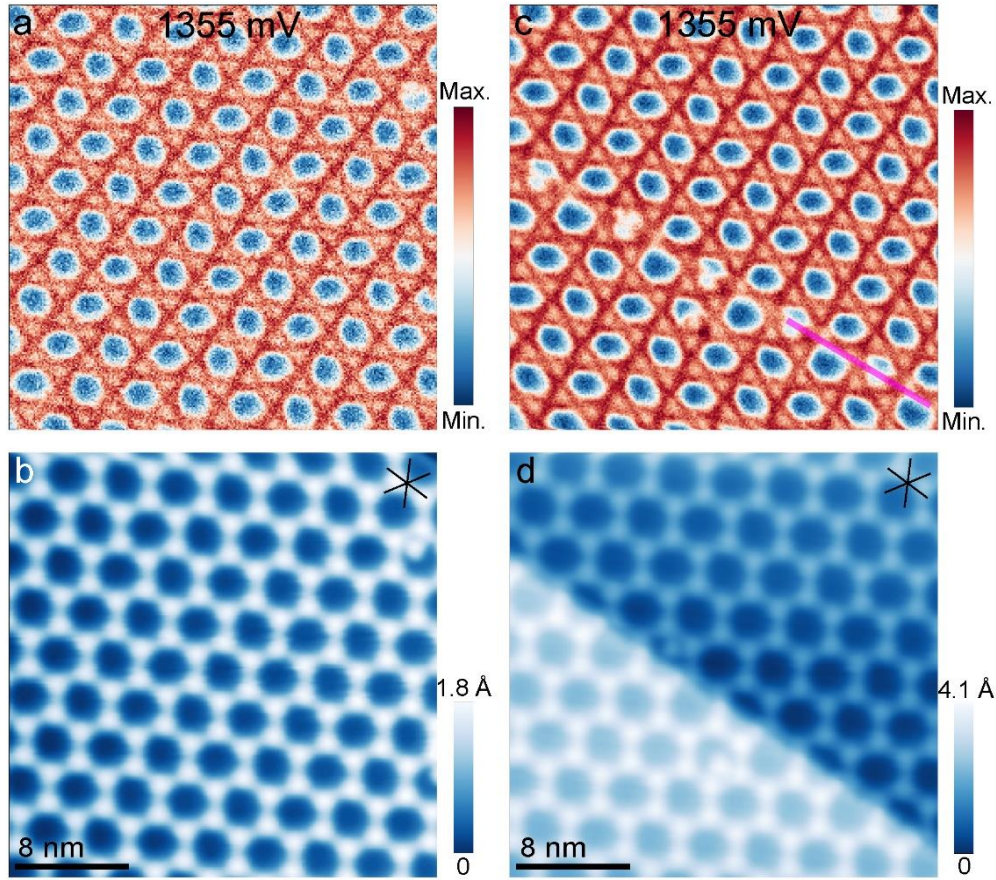

**Supplementary Figure 10 | Larger-scale images of the Kagome electronic grid.** (a) Larger-scale (30 nm × 30 nm) d$I$/d$V$ map of real-space Kagome electronic grid at 1355 mV, distributed homogeneously on the entire Ag-GDY network. Set point: $I_t$ = 50 pA, $U_m$ = 20 mV. (b) Corresponding STM image of (a). Set point: $I_t$ = 50 pA, $U_b$ = 1.355V. (c) Kagome electronic grid, which is continuous at atomically straight and monoatomic-height step at 1355 mV. Set point: $I_t$ = 50 pA, $U_m$ = 20 mV. (d) Corresponding STM image of (c). Set point: $I_t$ = 50 pA, $U_b$ = 1.355V. High-symmetry directions of the Ag(111)/mica surface are indicated.

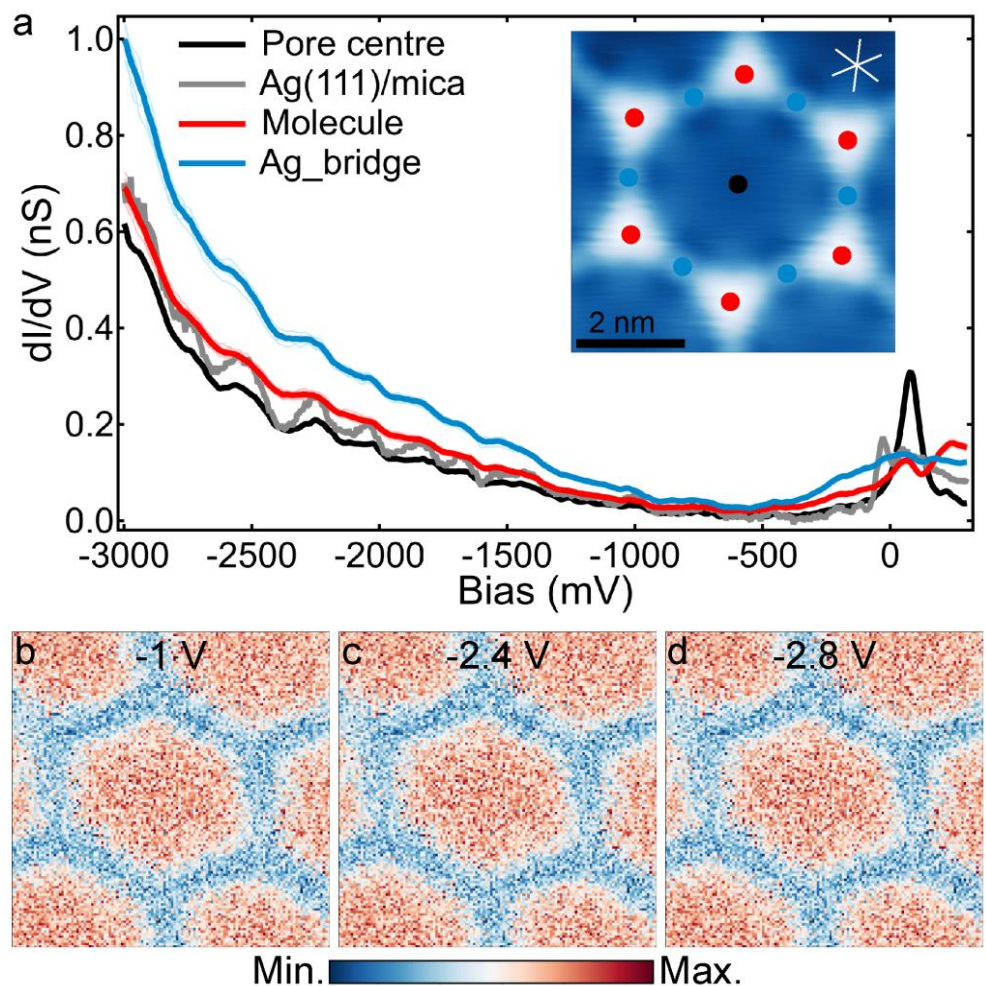

**Supplementary Figure 11 | d$I$/d$V$ point spectra of Ag-GDY network at occupied regime.** (a) Site-dependent d$I$/d$V$ spectra of a single hexagonal constituent of Ag-GDY network. The LDOS is featureless below Fermi energy. Set point: $I_t$ = 30 pA, $U_b$ = 0.30 V, $U_m$ = 20 mV. Inset, $I_t$ = 30 pA, $U_b$ = 300 mV. (b-d) d$I$/d$V$ maps of a single pore at selected bias voltages. Set point: $I_t$ = 50 pA, $U_m$ = 20 mV.

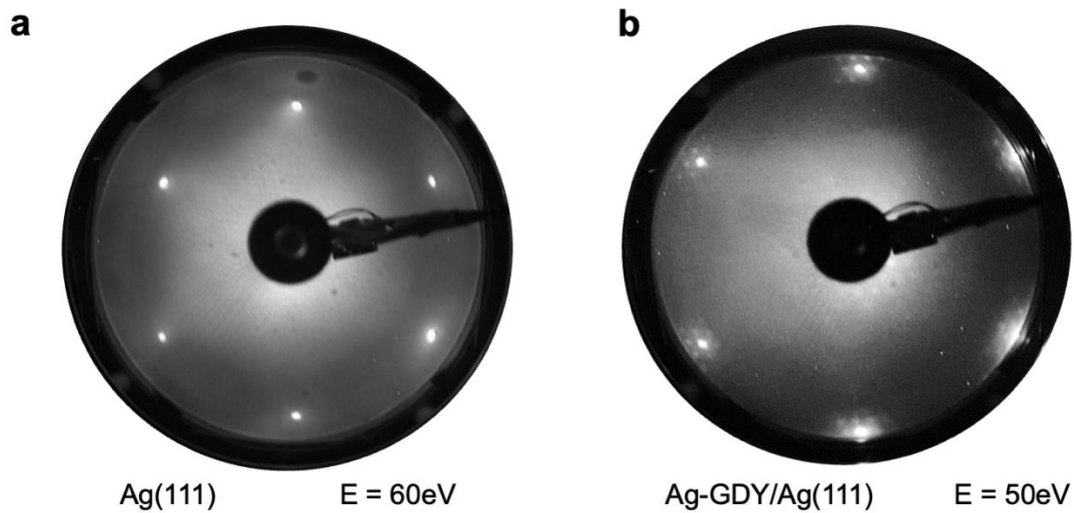

**Supplementary Figure 12 | LEED pattern comparison between pristine Ag(111) and Ag-GDY network fully covered Ag(111) surface during ARPES characterization.** (a) LEED pattern obtained for the pristine Ag(111) surface at 60 eV. (b) LEED pattern corresponding to the Ag-GDY network fully covered Ag(111) surface obtained at 50 eV. The diffraction pattern is identical to the one shown in Fig. 2b in the main text.

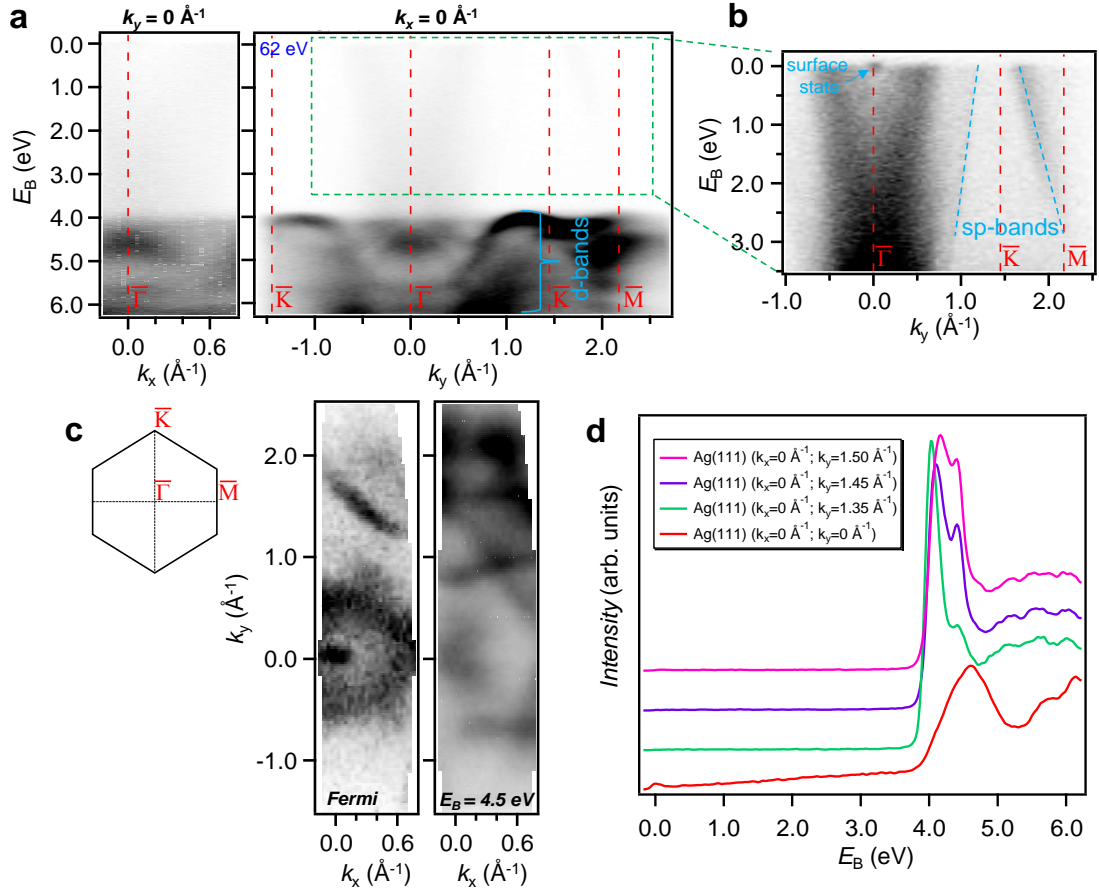

**Supplementary Figure 13 | Electronic structure of the Ag(111) crystal.** (a) ARPES band structure ($E$ vs $k_x$ at $k_y$=0 Å$^{-1}$) (left) and ($E$ vs $k_y$ at $k_x$=0 Å$^{-1}$) (right), corresponding to the direction parallel to the [11$\bar{2}$] and [1$\bar{1}$0] respectively. The characteristic silver *d*-bands appear below 4 eV. (b) A closer inspection into the valence band reveals the characteristic parabolic Ag surface state at the $\overline{\Gamma}$ point and the dispersive character of the *sp*-bands, as they raise and cross the Fermi level (highlighted by a side dashed light blue line). (c) Fermi surface map ($k_x$ vs $k_y$ at $E$=0 eV) (left) and at $E$=4.5 eV (right), exhibiting the isotropic Ag surface state close to the $\overline{\Gamma}$ point, the dispersive dominant *sp*-bands close to the $\overline{K}$ point and the hexagonal shape of the *d*-bands. At 62 eV photon energy, an additional spectral intensity is captured at $\overline{\Gamma}$ stemming from the projection of bulk bands. (d) Energy distribution curves (EDCs) extracted at different $k_y$ values along the $\overline{\Gamma K}$ direction. They clearly portray the Ag surface state and intensity from the projected bulk bands at $\overline{\Gamma}$ along with the highly intense onset of the *d*-bands beyond 4 eV.

S17

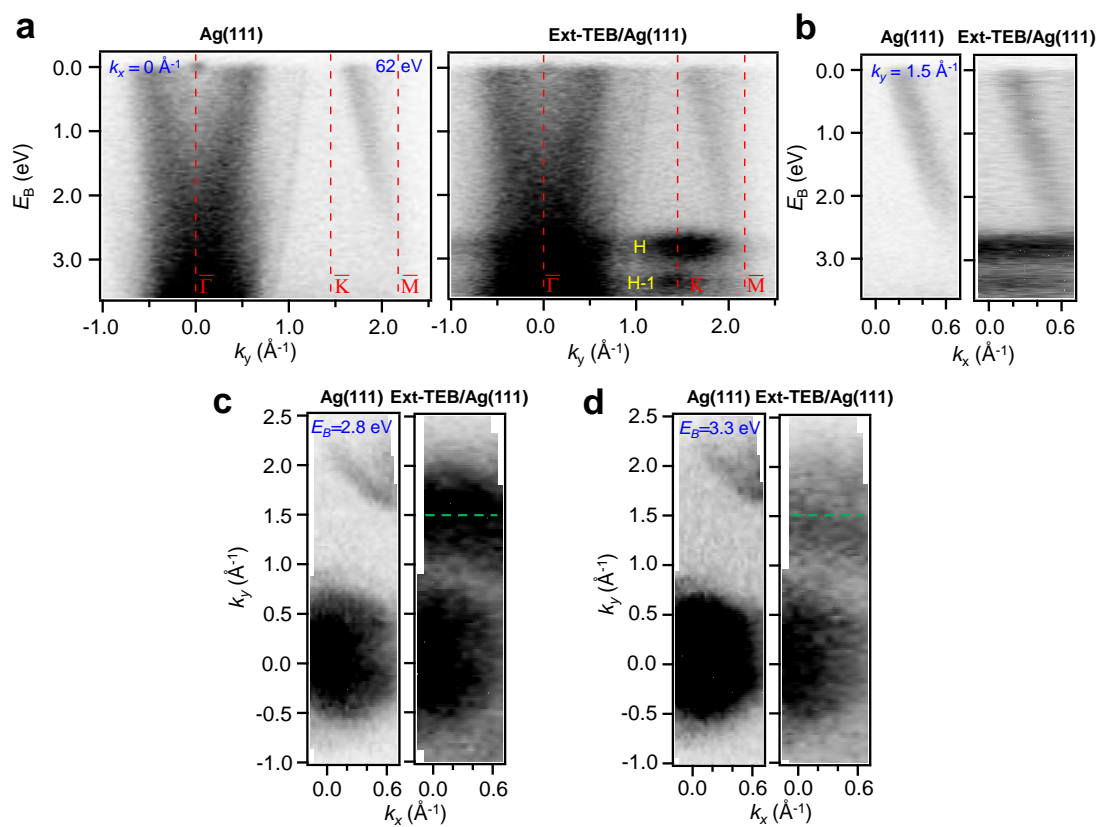

**Supplementary Figure 14 | ARPES band structure comparison between pristine Ag(111) and intact Ext-TEB multilayer on Ag(111).** (a) ARPES band structure ($E$ vs $k_y$ at $k_x=0$ Å$^{-1}$) along the [1$\bar{1}$0] direction for the pristine Ag (left) and intact Ext-TEB multilayer (right). Localized frontier molecular orbitals named as HOMO (H) and HOMO-1 (H-1) appear at the characteristic wave-vector of $k_y=1.5$ Å$^{-1}$. (b) The orthogonal ARPES band structure ($E$ vs $k_x$ at $k_y=1.5$ Å$^{-1}$) clearly evidences the lack of dispersion of the H and H-1. (c) Isoenergetic cuts ($k_x$ vs $k_y$) at the H level ($E=2.8$ eV) and (d) at the H-1 level ($E=3.3$ eV) as compared to the pristine Ag(111) case.

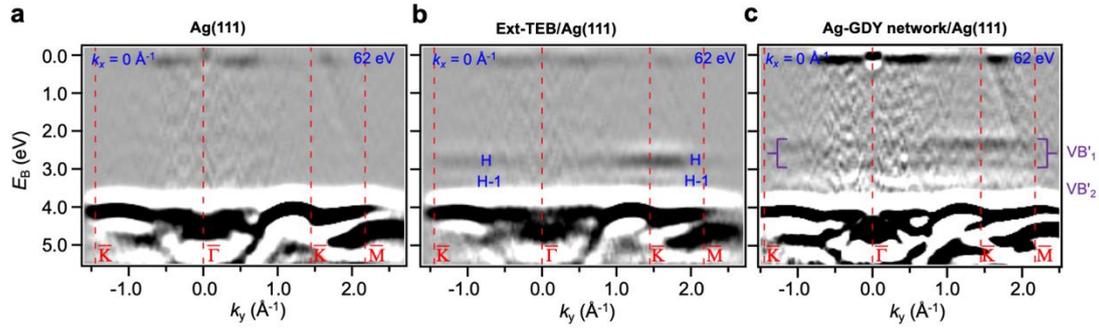

**Supplementary Figure 15 | ARPES band structure in second derivative comparing Ag(111), Ext-TEB/Ag(111) and Ag-GDY network/Ag(111).** ARPES band structure ($E$ vs $k_y$ at $k_x$=0 Å$^{-1}$) along the $[1\bar{1}0]$ direction, comparing the pristine Ag(111) (a), intact Ext-TEB multilayer (b) and Ag-GDY network (c). The localized molecular orbitals (denoted as H and H-1) in (b) evolve into valence bands ($VB'_1$ and $VB'_2$) upon the formation of the Ag-GDY network. The wide bandwidth of $VB'_1$ evidences the delocalized electronic states in the Ag-GDY network. No appreciable changes are observed for the substrate $d$-bands, which display better resolution in (c) due to the lower substrate temperature during measurement ($T_{sub}$ = 60 K) in comparison to the RT measurements in (a) and (b).

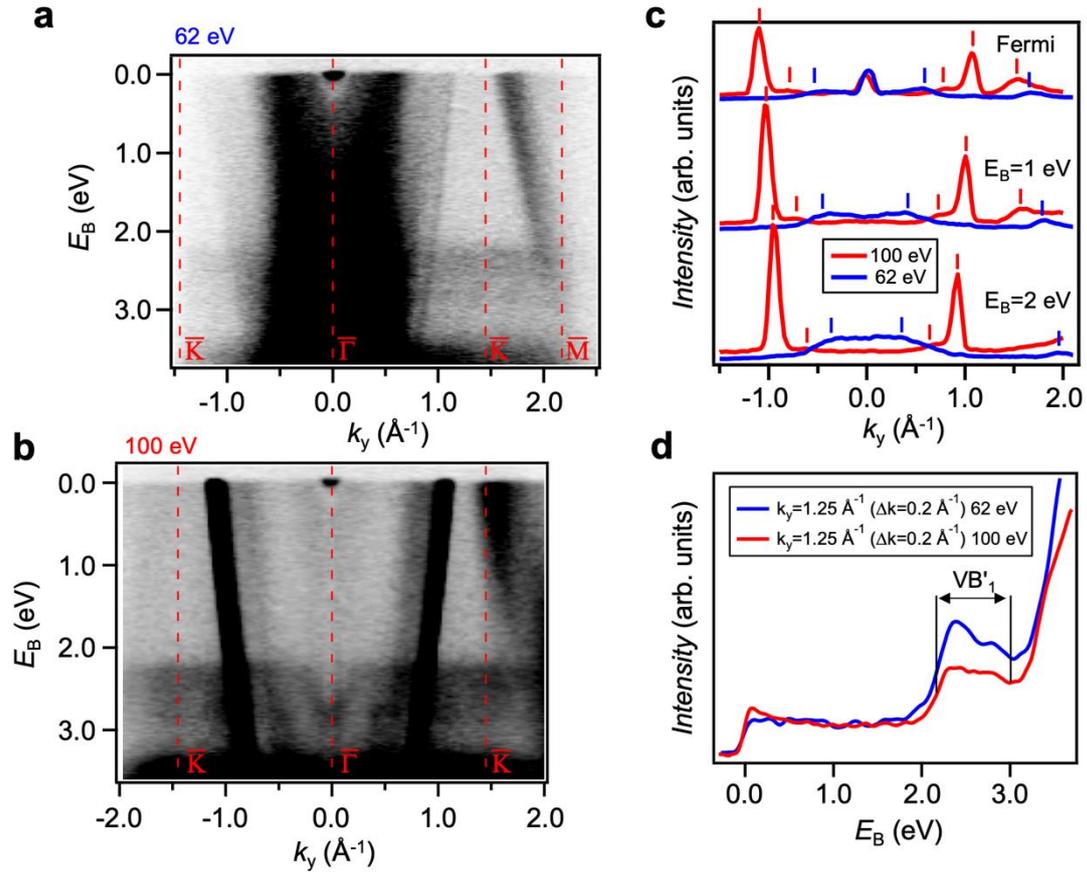

**Supplementary Figure 16 | Ag-GDY ARPES band structure dependence with photon energy.** (a, b) ARPES band structures ($E$ vs $k_y$ at $k_x$=0 Å$^{-1}$) of the Ag-GDY network along the [1$\bar{1}$0] direction of Ag(111) measured at 62 eV (a) and 100eV (b) photon energies, respectively. (c) Constant energy curves at different binding energies (Fermi level, 1 eV and 2 eV) extracted from the ARPES spectra in (a) and (b). Clearly the Ag $sp$-bands change their position with different photon energies. (d) Energy distribution curves obtained at $k_y$=1.25 Å$^{-1}$ from (a) and (b) show that the $VB'_1$ and $VB'_2$ features remain unchanged with two different photon energies, evidencing their 2D nature.

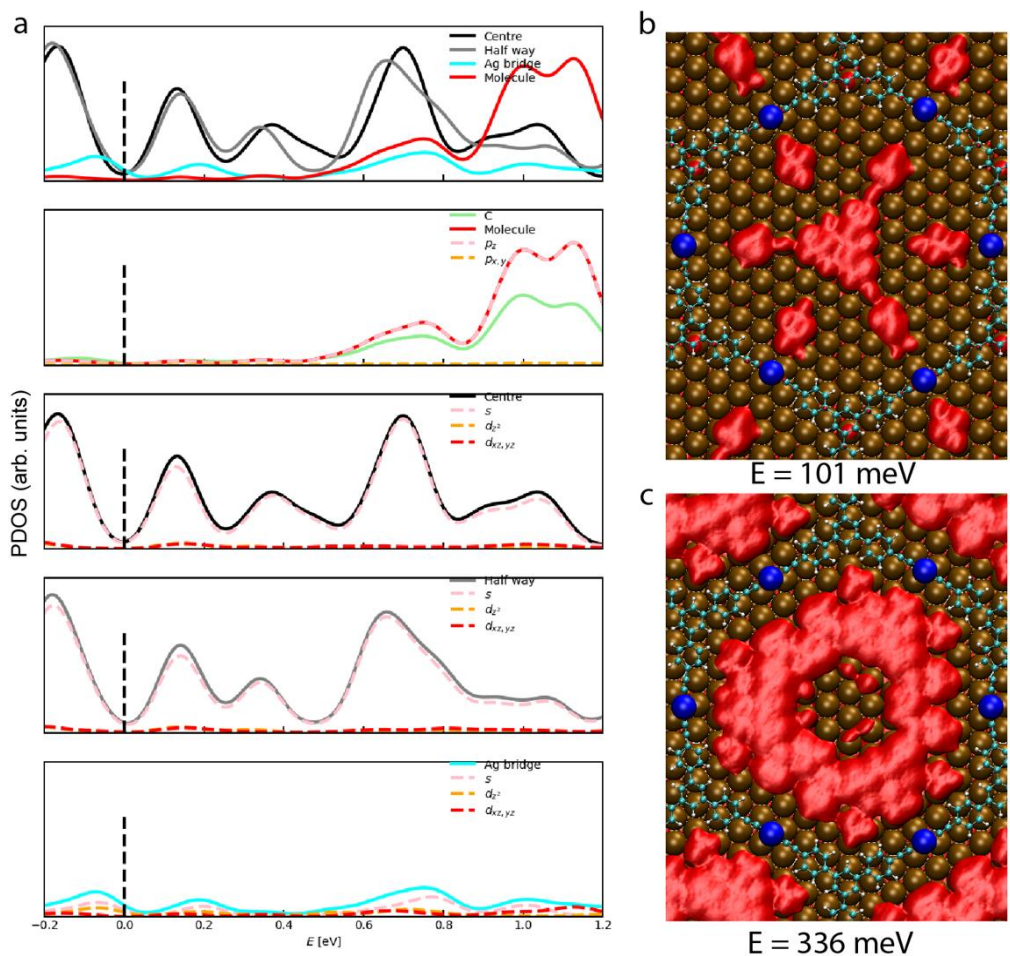

**Supplementary Figure 17 | DFT-calculated nanopore-confined surface states.** (a) Site-specific projected density of states of Ag-GDY network adsorbed on the four-layer Ag(111) slab. (b,c) Isosurfaces of orbital density of the whole system with two selected energies corresponding to n=1,2 of the nanopore-confined surface state.

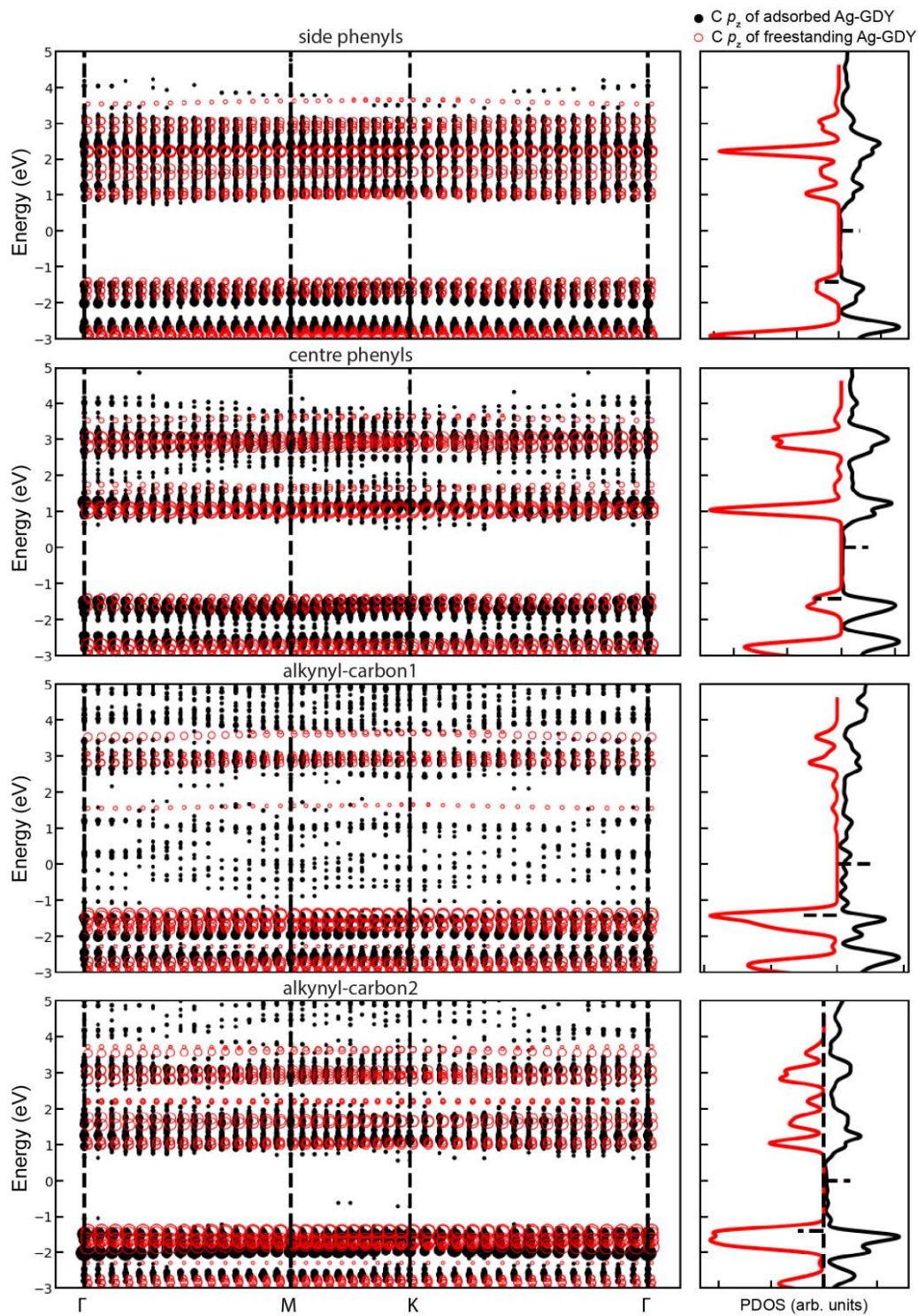

**Supplementary Figure 18 | Projected band structure and DOS of carbon $p_z$ component from phenyl rings and alkynes in the Ag-GDY network.** Red circles and black dots correspond to the free-standing and the adsorbed Ag-GDY network, respectively. The Fermi levels of two systems are shifted to achieve good alignment between the band gaps. Alkynyl-carbon1/2 represents the carbon atom in the alkynyl group connecting to the phenyl-ring (or substrate).

Hybridization and charge transfer effects are well-known effects for organometallic or metal-organic nanostructures fabricated on noble metal surfaces[2]. Note that the degree of hybridization (mild or strong) depends on the noble metal crystal used[3] as well as organic system under investigation.

In our case, the band structure for the free-standing Ag-GDY sheet is moderately altered when adsorbed on the Ag(111) substrate (see Figures S18-20). This can be recognized by comparing the free-standing Ag-GDY bands (in red or blue) with the ones corresponding to the whole system (in black or purple). Due to the presence of charge transfer from the substrate to the Ag-GDY layer, PDOS peaks of freestanding and adsorbed sheets are aligned for comparison. Both peak positions as well as their widths do not show significant variation, which clearly indicates a mild hybridization degree[3].

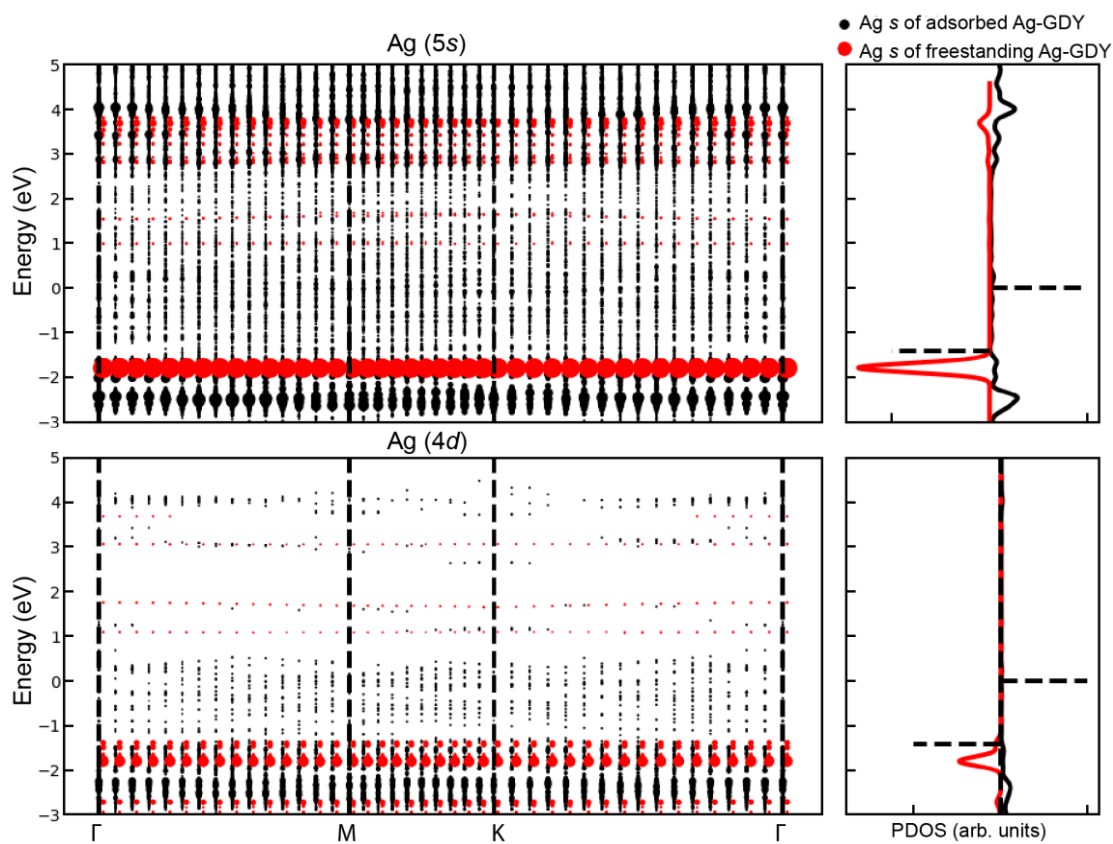

**Supplementary Figure 19 | Projected band structure and DOS of Ag(5s) and Ag(4d) components from alkynyl-Ag atom in the Ag-GDY network.** Red and black dots correspond to the free-standing and the adsorbed network, respectively. The Fermi levels of two systems are shifted following the value according to Supplementary Fig. 18.

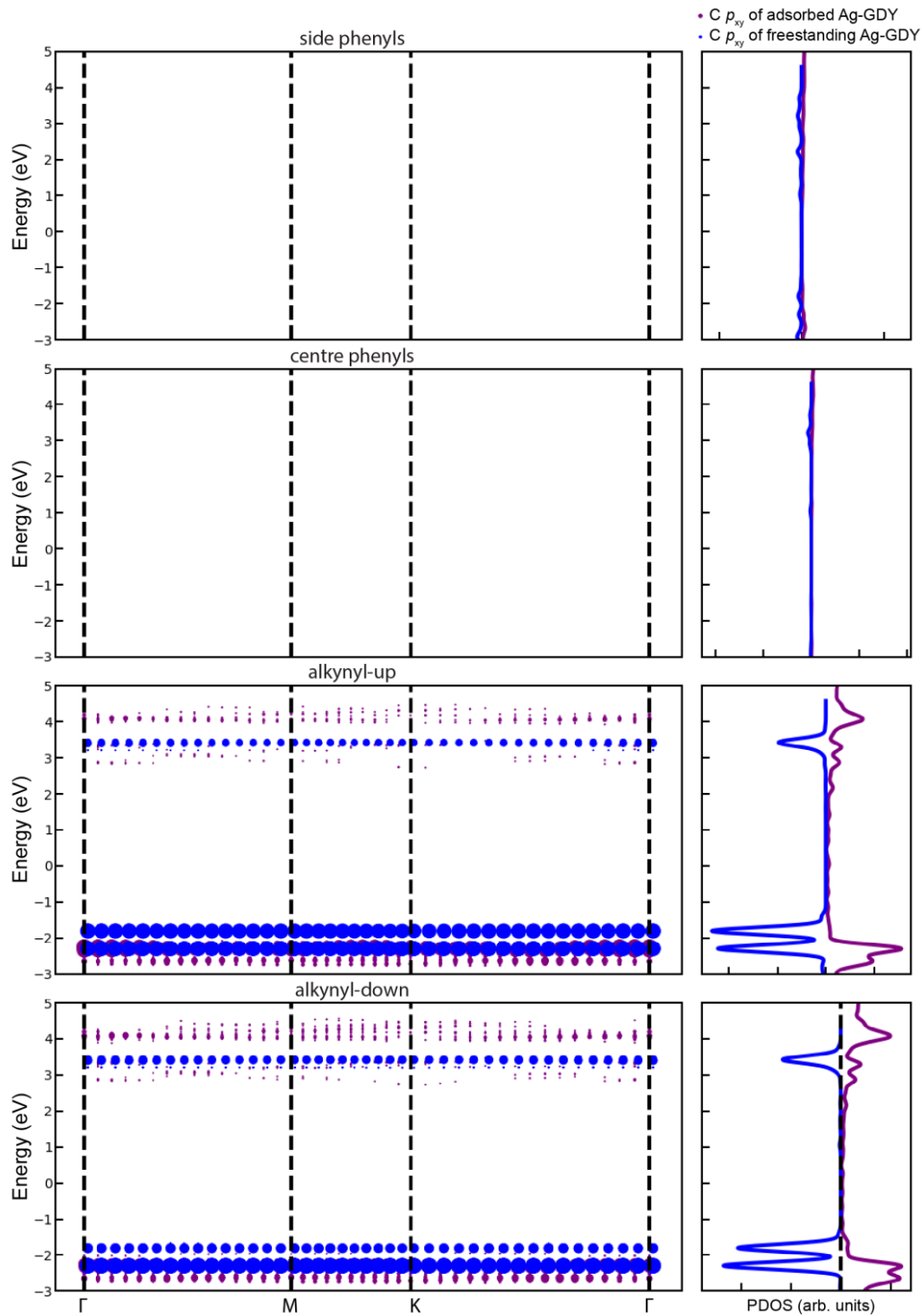

**Supplementary Figure 20 | Projected band structure and DOS of carbon in-plane p components from phenyl rings and alkynes in the Ag-GDY network.** Blue and violet dots correspond to the free-standing and the adsorbed network, respectively. The Fermi levels of two systems are shifted following the value according to Supplementary Fig. 18.